\documentclass[reprint,superscriptaddress,nofootinbib,amsmath,amssymb,aps,prd,]{revtex4-2}

\usepackage{amsmath}        
\usepackage{amsfonts}       
\usepackage{bbm}            

\usepackage{graphicx}       
\usepackage{float}          
\usepackage{placeins}		

\usepackage{multirow}       
\usepackage{booktabs}

\usepackage{url}            
\usepackage{acro}			

\usepackage{xcolor}         
\usepackage{textgreek}		
\usepackage{siunitx}		

\DeclareAcroArticle{definite}{the}
\NewAcroCommand\dac{m}{\acrodefinite\UseAcroTemplate{first}{#1}}
\NewAcroCommand\Dac{m}{\acroupper\acrodefinite\UseAcroTemplate{first}{#1}}
\NewAcroCommand\dacp{m}{\acroplural\acrodefinite\UseAcroTemplate{first}{#1}}
\NewAcroCommand\Dacp{m}{\acroupper\acroplural\acrodefinite\UseAcroTemplate{first}{#1}}
\NewAcroCommand\dacs{m}{\acrodefinite\UseAcroTemplate{short}{#1}}
\NewAcroCommand\Dacs{m}{\acroupper\acrodefinite\UseAcroTemplate{short}{#1}}
\NewAcroCommand\dacsp{m}{\acroplural\acrodefinite\UseAcroTemplate{short}{#1}}
\NewAcroCommand\Dacsp{m}{\acroupper\acroplural\acrodefinite\UseAcroTemplate{short}{#1}}
\NewAcroCommand\daca{m}{\acrodefinite\UseAcroTemplate{alt}{#1}}
\NewAcroCommand\Daca{m}{\acroupper\acrodefinite\UseAcroTemplate{alt}{#1}}
\NewAcroCommand\dacap{m}{\acroplural\acrodefinite\UseAcroTemplate{alt}{#1}}
\NewAcroCommand\Dacap{m}{\acroupper\acroplural\acrodefinite\UseAcroTemplate{alt}{#1}}
\NewAcroCommand\dacl{m}{\acrodefinite\UseAcroTemplate{long}{#1}}
\NewAcroCommand\Dacl{m}{\acroupper\acrodefinite\UseAcroTemplate{long}{#1}}
\NewAcroCommand\daclp{m}{\acroplural\acrodefinite\UseAcroTemplate{long}{#1}}
\NewAcroCommand\Daclp{m}{\acroupper\acroplural\acrodefinite\UseAcroTemplate{long}{#1}}
\NewAcroCommand\dacf{m}{\acrodefinite\acrofull\UseAcroTemplate{first}{#1}}
\NewAcroCommand\Dacf{m}{\acroupper\acrodefinite\acrofull\UseAcroTemplate{first}{#1}}
\NewAcroCommand\dacfp{m}{\acroplural\acrodefinite\acrofull\UseAcroTemplate{first}{#1}}
\NewAcroCommand\Dacfp{m}{\acroupper\acroplural\acrodefinite\acrofull\UseAcroTemplate{first}{#1}}
\DeclareAcronym{DM}{
	short = DM,
	long = dark matter
}
\DeclareAcronym{MW}{
	short = MW,
	long = Milky Way
}
\DeclareAcronym{GAMBIT}{
	short = GAMBIT,
	long = Global and Modular Beyond-the-Standard-Model Inference Tool,
	cite = GAMBIT:2017yxo,
	short-definite = \nospace%
}
\DeclareAcronym{DD}{
	short = DD,
	long = direct detection
}
\DeclareAcronym{SI}{
	short = SI,
	long = spin independent
}
\DeclareAcronym{SD}{
	short = SD,
	long = spin dependent
}
\DeclareAcronym{NREO}{
	short = NREO,
	long = non-relativistic effective operator
}
\DeclareAcronym{DE}{
	short = DE,
	long = differential evolution
}
\DeclareAcronym{SM}{
	short = SM,
	long = Standard Model
}
\DeclareAcronym{GFFI}{
	short = GFFI,
	long = Generalized Form Factor Integral
}
\DeclareAcronym{MCMC}{
	short = MCMC,
	long = Markov Chain Monte Carlo
}

\ifdefined\qty\else
	\ifdefined\NewCommandCopy 
		\NewCommandCopy\qty\SI 
	\else
		\NewDocumentCommand\qty{O{}mm}{\SI[#1]{#2}{#3}}
	\fi
\fi
\ifdefined\unit\else
	\ifdefined\NewCommandCopy 
		\NewCommandCopy\unit\si 
	\else
		\NewDocumentCommand\unit{O{}m}{\si[#1]{#2}}
	\fi
\fi
\ifdefined\qtyproduct\else
	\ifdefined\DeclareCommandCopy 
		\DeclareCommandCopy\SI\qtyproduct 
	\else
		\DeclareDocumentCommand\SI{O{}mm}{\qtyproduct[#1]{#2}{#3}}
	\fi
\fi
\DeclareSIUnit\litre{L}
\DeclareSIUnit\liter{\litre}

\newcommand{\diff}{\mathop{} \! \mathrm{d}}

\newcommand{\bbar}{\overline{b}}
\newcommand{\bb}{b\bbar}

\newcommand{\Wp}{W^{+}}
\newcommand{\Wm}{W^{-}}
\newcommand{\WW}{\Wp\Wm}

\newcommand{\taup}{\tau^{+}}
\newcommand{\taum}{\tau^{-}}
\newcommand{\tautau}{\taup\taum}

\newcommand{\Op}[1]{\widehat{\mathcal{O}}_{#1}}
\newcommand{\couple}[2][\tau]{c^{#1}_{#2}}

\newcommand{\vecOp}[1]{\mathbf{\widehat{#1}}}

\newcommand{\N}{\mathrm{N}}

\newcommand{\hermo}{\mathbbm{1}_{\chi\N}}
\newcommand{\hermq}{\vecOp{q}}
\newcommand{\vel}{\vecOp{v}}
\newcommand{\hermv}{{\vel}^{\perp}}
\newcommand{\hermS}[1]{\vecOp{S}_{#1}}

\newcommand{\mN}{m_\N}

\newcommand{\vtangsq}{{\vel}^{\perp{}2}_{\mathrm{T}}}
\newcommand{\cdms}[1][n]{CDMSlite%
\ifx c#1
		~\cite{SuperCDMS:2015eex}
	\else
		\ifx n#1
		\else 
			\PackageError{Experiment Macros}{Unclear citation instruction}{Either use n for no citeations, or c for citations}
		\fi
	\fi
}
\newcommand{\cresstII}[1][n]{CRESST-II%
\ifx c#1
		~\cite{CRESST:2015txj}
	\else
		\ifx n#1
		\else 
			\PackageError{Experiment Macros}{Unclear citation instruction}{Either use n for no citeations, or c for citations}
		\fi
	\fi
}
\newcommand{\cresstIII}[1][n]{CRESST-III%
\ifx c#1
		~\cite{CRESST:2019jnq}
	\else
		\ifx n#1
		\else 
			\PackageError{Experiment Macros}{Unclear citation instruction}{Either use n for no citeations, or c for citations}
		\fi
	\fi
}
\newcommand{\darksideSTwo}[1][n]{DarkSide-50-S2-Only%
\ifx c#1
		~\cite{DarkSide:2018bpj}
	\else
		\ifx n#1
		\else 
			\PackageError{Experiment Macros}{Unclear citation instruction}{Either use n for no citeations, or c for citations}
		\fi
	\fi
}
\newcommand{\darkside}[1][n]{DarkSide-50%
\ifx c#1
		~\cite{DarkSide:2018kuk}
	\else
		\ifx n#1
		\else 
			\PackageError{Experiment Macros}{Unclear citation instruction}{Either use n for no citeations, or c for citations}
		\fi
	\fi
}
\newcommand{\lux}[1][n]{LUX%
\ifx c#1
		~\cite{LUX:2016ggv}
	\else
	\ifx n#1
	\else 
			\PackageError{Experiment Macros}{Unclear citation instruction}{Either use n for no citeations, or c for citations}
		\fi
	\fi
}
\newcommand{\lz}[1][n]{LZ%
\ifx c#1
		~\cite{LZ:2022lsv}
	\else
	\ifx n#1
	\else 
			\PackageError{Experiment Macros}{Unclear citation instruction}{Either use n for no citeations, or c for citations}
		\fi
	\fi
}
\newcommand{\pandaII}[1][n]{PandaX-II%
\ifx c#1
		~\cite{PandaX-II:2016vec,PandaX-II:2017hlx}
	\else
		\ifx n#1
		\else 
			\PackageError{Experiment Macros}{Unclear citation instruction}{Either use n for no citeations, or c for citations}
		\fi
	\fi
}
\newcommand{\pandaFourT}[1][n]{PandaX-4T%
\ifx c#1
		~\cite{PandaX-4T:2021bab}
	\else
		\ifx n#1
		\else 
			\PackageError{Experiment Macros}{Unclear citation instruction}{Either use n for no citeations, or c for citations}
		\fi
	\fi
}
\newcommand{\picoTwoL}[1][n]{PICO-2L%
\ifx c#1
		~\cite{PICO:2016kso}
	\else
		\ifx n#1
		\else 
			\PackageError{Experiment Macros}{Unclear citation instruction}{Either use n for no citeations, or c for citations}
		\fi
	\fi
}
\newcommand{\pico}[1][n]{PICO-60%
\ifx c#1
		~\cite{PICO:2015pux,PICO:2017tgi,PICO:2019vsc}
	\else
		\ifx n#1
		\else 
			\PackageError{Experiment Macros}{Unclear citation instruction}{Either use n for no citeations, or c for citations}
		\fi
	\fi
}
\newcommand{\simple}[1][n]{SIMPLE%
\ifx c#1
		~\cite{SIMPLE:2014pun}
	\else
		\ifx n#1
		\else 
			\PackageError{Experiment Macros}{Unclear citation instruction}{Either use n for no citeations, or c for citations}
		\fi
	\fi
}
\newcommand{\picoFive}[1][n]{PICO-500%
\ifx c#1
		~\cite{Fallows:2017}
	\else
		\ifx n#1
		\else 
			\PackageError{Experiment Macros}{Unclear citation instruction}{Either use n for no citeations, or c for citations}
		\fi
	\fi
}
\newcommand{\xenonOneH}[1][n]{XENON100%
\ifx c#1
		~\cite{XENON100:2012itz}
	\else
		\ifx n#1
		\else 
			\PackageError{Experiment Macros}{Unclear citation instruction}{Either use n for no citeations, or c for citations}
		\fi
	\fi
}
\newcommand{\xenonOneT}[1][n]{XENON1T%
\ifx c#1
		~\cite{XENON:2018voc}
	\else
		\ifx n#1
		\else 
			\PackageError{Experiment Macros}{Unclear citation instruction}{Either use n for no citeations, or c for citations}
		\fi
	\fi
}
\newcommand{\antares}[1][n]{ANTARES%
\ifx c#1
		~\cite{Poire:2022}
	\else
		\ifx n#1
		\else 
			\PackageError{Experiment Macros}{Unclear citation instruction}{Either use n for no citeations, or c for citations}
		\fi
	\fi
}
\newcommand{\ic}[1][n]{IceCube%
\ifx c#1
		~\cite{IceCube:2012ugg,IceCube:2012fvn,IceCube:2016yoy,IceCube:2022wxw}
	\else
		\ifx n#1
		\else 
			\PackageError{Experiment Macros}{Unclear citation instruction}{Either use n for no citeations, or c for citations}
		\fi
	\fi
}
\newcommand{\deepcore}[1][n]{DeepCore%
\ifx c#1
		~\cite{IceCube:2021xzo}
	\else
		\ifx n#1
		\else 
			\PackageError{Experiment Macros}{Unclear citation instruction}{Either use n for no citeations, or c for citations}
		\fi
	\fi
}
\newcommand{\superk}[1][n]{Super-Kamiokande%
\ifx c#1
		~\cite{Super-Kamiokande:2015xms}
	\else
		\ifx n#1
		\else 
			\PackageError{Experiment Macros}{Unclear citation instruction}{Either use n for no citeations, or c for citations}
		\fi
	\fi
}
\newcommand{\darwin}[1][n]{DARWIN%
\ifx c#1
		~\cite{Schumann:2015cpa}
	\else
		\ifx n#1
		\else 
			\PackageError{Experiment Macros}{Unclear citation instruction}{Either use n for no citeations, or c for citations}
		\fi
	\fi
}

\newcommand{\allDD}[1][n]{\cdms[#1], \cresstII[#1], \cresstIII[#1], \darkside[#1], \darksideSTwo[#1], \lux[#1], \lz[#1], \pandaII[#1], \pandaFourT[#1], \picoTwoL[#1], \pico[#1], \simple[#1], \xenonOneH[#1], and \xenonOneT[#1]}

\usepackage[colorlinks]{hyperref}
\hypersetup{
	colorlinks,
	linkcolor={red!50!black},
	citecolor={blue!50!black},
	urlcolor={blue!80!black}
}
\usepackage{orcidlink}		

\begin{document}

\title{A Global Fit of Non-Relativistic Effective Dark Matter Operators Including Solar Neutrinos}
\newcommand{\QU}{Department of Physics, Engineering Physics, and Astronomy, Queen's University, Bader Lane, Kingston, Canada}
\newcommand{\MI}{Arthur B. McDonald Canadian Astroparticle Physics Research Institute, Queen's University, Bader Lane, Kingston, Canada}
\newcommand{\PI}{Perimeter Institute for Theoretical Physics, Caroline Street North, Waterloo, Canada}
\newcommand{\QB}{Quantum Brilliance Pty Ltd, The Australian National University, Daley Road, Acton ACT 2601, Australia}

\affiliation{\QU}
\affiliation{\MI}
\affiliation{\PI}
\affiliation{\QB}

\author{N.\,P.~Avis~Kozar\,\orcidlink{0000-0003-4875-6782}}\affiliation{\QU}\affiliation{\MI}
\author{Pat Scott\,\orcidlink{0000-0002-3151-3701}} \affiliation{\QB}
\author{A.\,C.~Vincent\,\orcidlink{0000-0003-3872-0743}}\affiliation{\QU}\affiliation{\MI}\affiliation{\PI}
\date{\today}

\begin{abstract}
	We perform a global fit of \acl{DM} interactions with nucleons using a \acl{NREO} description, considering both \acl{DD} and neutrino data.
	We examine the impact of combining the \acl{DD} experiments \allDD[] along with neutrino data from \ic[], \antares[], \deepcore[], and \superk[].
	While current neutrino telescope data lead to increased sensitivity compared to underground nuclear scattering experiments for \acl{DM} masses above \qty{100}{\GeV}, our future projections show that the next generation of underground experiments will significantly outpace solar searches for most dark matter-nucleon elastic scattering interactions.
\end{abstract}

\maketitle
\acresetall{}

\section{Introduction}\label{sec:Intro}

	Observations across a range of cosmic scales, from galactic rotation curves~\cite{Persic:1995ru} to the angular power spectrum of the cosmic microwave background~\cite{Planck:2018vyg}, point to a large component of weakly-interacting, cold \ac{DM}.
	The nature of \ac{DM} remains unknown.
	If it is a weakly-interacting particle in the \( \sim \) \unit{\GeV}--\unit{\TeV} range, it could leave an imprint in \ac{DD} experiments, which search for energy deposition from \ac{DM} elastically scattering with target nuclei.
	Thus far, no detection has been confirmed, and increasingly strong limits have been placed on \ac{DM} interacting with nucleons via a constant \ac{SD} or \ac{SI} coupling, respectively parametrized via their cross section with nucleons \( \sigma_\text{SI} \) and \( \sigma_\text{SD} \).
	These limits span many orders of magnitude in mass and cross section; see e.g.~\cite{Cerdeno:2010jj,Lisanti:2016jxe,Schumann:2019eaa}.
 
	Possible microphysical interactions extend well beyond \( \sigma_{\rm SI} \) and \( \sigma_{\rm SD} \).
	Over the past decade, a formalism has been developed to parametrise the set of non-relativistic effective interactions in the low-energy limit of a generic dark sector model~\cite{Fan:2010gt,Kumar:2013iva,Fitzpatrick:2012ix}.
	By decomposing possible interaction Hamiltonian terms into products of available Hermitian operators that may couple \ac{DM} to nucleons, one finds cross sections that scale with powers of the relative \ac{DM}-nucleus velocity, transferred momentum, or particle spins, which in turn give rise to different phenomenology.
	The 14 operators that make up the \ac{NREO} formalism have now become standard; see e.g.~\cite{Dent:2015zpa} for a full description, including how they may arise from a relativistic theory.

	Any interaction between \ac{DM} and nuclei in the lab also implies interactions in celestial bodies such as stars.
	Should \ac{DM} have some novel interaction with baryonic matter, then through kinematic scattering with nuclei in the Sun the \ac{DM} could become gravitationally bound to it.
	This may lead to a neutrino signal if the \ac{DM} can self-annihilate to \ac{SM} final states.
	Extensive searches for such a signal have been performed by neutrino telescopes including \superk[c], ANTARES~\cite{ANTARES:2016xuh,Poire:2021vbr} and IceCube~\cite{IceCube:2012ugg,IceCube:2012fvn,IceCube:2016yoy,Arguelles:2019jfx,IceCube:2022wxw}.
	These are presented as constraints on the \ac{DM}-nucleus scattering cross section \( \sigma_\text{SI} \), \( \sigma_\text{SD} \) since in a steady state, the annihilation rate is equal to the capture rate, proportional to \(\sigma_{\chi n}\).
	It is important to note that these constraints are strongly dependent on the annihilation channel, each of which yields a specific spectrum and flux of neutrinos.

	The Sun and underground experiments probe different kinematic regimes: \ac{DD} is by design most sensitive to high-velocity and high momentum-transfer interactions, and benefits from the \(A^2\) coherent enhancement of scattering with heavy nuclei; the Sun, primarily composed of hydrogen, presents a large target for spin-dependent scattering, and can more readily sample the low-velocity tail of the dark matter phase space distribution.
	These two methods are thus complementary in the types of interactions they may see.

	Previous work has examined the effect of non-constant \ac{DM}-nucleus interactions on the capture rate of dark matter in the Sun.
	Ref.~\cite{Vincent:2015gqa} obtained capture rates for generic spin-dependent and independent interactions that scale with the relative velocity and momentum transfer.
	Ref.~\cite{Catena:2015uha} computed cross sections and capture rates in the context of the \ac{NREO} formalism, and found parametrizations for the effective nuclear response (or \textit{form factor}) for the most abundant elements in the Sun.
	In Ref.~\cite{Catena:2015iea}, these were used to place constraints on 28 different coupling constants using one year of IceCube 79-string data, as well as reported data from \superk[].
	Ref~\cite{Catena:2018vzc} further constrained the effective theory of \textit{inelastic} effective interactions, and complementarity between experiments in the \ac{NREO} framework was examined in Ref.~\cite{Brenner:2022qku}.

	In contrast with~\cite{Brenner:2022qku} where the authors looked to emphasize potentially incorrect conclusions drawn from the comparison of specific models with individual interaction limits, here we wish to investigate a global analysis of \acp{NREO}.
 	As a result we do not pick specific high energy theories corresponding to a subset of interaction types, but rather investigate the relative contributions to each interaction from the experiments considered.
	While this work included visual comparison of solar neutrino constraints with direct detection results, no self-consistent combination of likelihoods from both sectors has been considered in the context of \ac{NREO} searches. That is to say, self-consistently including common nuisance factors in the computation of likelihoods across all experiments and between direct detection and solar neutrino constraints.
	Further, neutrino data included in previous analyses has been limited to only a small fraction of the more than ten years over which IceCube has operated.
 
	In this work, we thus implement solar capture and direct detection calculations in the \ac{NREO} interaction formalism into \dac{GAMBIT}, and perform a global fit of this parameter space including possible nuisance parameters that both sectors may have in common.
	By combining likelihoods before reporting constraints, we ensure self-consistency, and may access parameter space that is not covered by simply overlaying constraints from individual experiments~\cite{AbdusSalam:2020rdj}
	
	We make use of the \ac{DD} experiments included in \dac{GAMBIT}.
	These include liquid noble gas detectors, super-heated fluorine experiments, and cryogenic crystal experiments: \allDD[c].
	The solar constraints come from \ic[c], \antares[c], \superk[c], and \deepcore[c] neutrino data.
	Of particular importance, IceCube has recently shown preliminary constraints on dark matter capture and annihilation in the Sun~\cite{IceCube:2022wxw} using 9 years of 86-string data, which we recast here.
	We will furthermore show projected sensitivities for the some of the most significant planned experiments: \darwin[c], \picoFive[c], and a combined set of planned future neutrino experiments.

	The \ac{DD} constraints largely make use of work developed in the context of prior GAMBIT publications.
	Specifically, Ref.~\cite{GAMBIT:2021rlp} included a combination of \ac{DD} and IC79 constraints on relativistic effective operators; we refer the reader to that reference for details on the \ac{DD} implementation.
	Here, we will rather focus on the novel implementation of the solar capture rates computed within the \ac{NREO} formalism, their interplay with \ac{DD} experiments, and the future of such combined searches for dark matter.

	This article is organized as follows: in Sec.~\ref{sec:Theory}, we review the \acp{NREO} formalism, and in Sec.~\ref{Sec:Capture} we show how capture of \ac{DM} in the Sun is computed.
	Sec.~\ref{sec:Methods} covers the details of the software implementation in the Capt'n General code, designed to calculate solar \ac{DM} capture rates using the \ac{NREO} formalism, and how Capt'n General interfaces with \dac{GAMBIT}.
	Sec.~\ref{sec:Results} contains our main results, and finally Sec.~\ref{sec:Conclusions} summarizes the work of this paper and future prospects to expand on this work.
	The \dac{GAMBIT} scan outputs and script for reproducing the plots found in this paper can be found in a Zenodo dataset~\cite{aviskozar:2023}.

\section{Non-Relativistic Effective Operator Dark Matter}\label{sec:Theory}

	We make use of \acp{NREO} following the convention from~\cite{Catena:2015uha}, which are developed from~\cite{Fitzpatrick:2012ix}.
	This work focuses on the use of \acp{NREO} as a method to describe elastic interactions between \ac{DM} from the \ac{MW} halo and nuclei in the Sun.
	The operators are constructed to describe \ac{DM} interactions mediated by a massive particle of spin-zero or spin-one.
	These effective operators are constructed from five Hermitian operators that obey Galilean invariance.
	These are the transferred momentum \( i\hermq \), the relative transverse velocity \( \hermv \), the nucleon spin \( \hermS{\N} \), the \ac{DM} spin \( \hermS{\chi} \), and the identity \( \hermo \) in the \ac{DM}-nucleon space:
	\begin{equation}
		\hermo \; \; , \; \; i\hermq \; \; , \; \; \hermv \; \; , \; \; \hermS{\chi} \; \; , \; \; \hermS{\N} \, .
		\label{eqn:HermitianOps}
	\end{equation}
	The quantity \( \hermv = \vel + \hermq/(2\mu_N) \) is defined using the relative velocity between the nucleon and incoming \ac{DM} particle \( \vel \), and is constructed such that \( \hermq \cdot \hermv = 0 \), ensuring linear independence.

	\begin{table*}[htb]
		\centering
		\begin{tabular}{l l}
			\toprule
			\( \Op{ 1 } = \hermo \)																										& \( \Op{ 9 }  = i \hermS{\chi} \cdot \left( \hermS{\N} \times \frac{\hermq}{\mN} \right) \)																\\
			\( \Op{ 3 } = i \hermS{\N} \cdot \left( \frac{\hermq}{\mN} \times \hermv \right) \)											& \( \Op{ 10 } = i \hermS{\N} \cdot \frac{\hermq}{\mN} \)																									\\
			\( \Op{ 4 } = \hermS{\chi} \cdot \hermS{\N} \)																				& \( \Op{ 11 } = i \hermS{\chi} \cdot \frac{\hermq}{\mN} \)																									\\
			\( \Op{ 5 } = i \hermS{\chi} \cdot \left( \frac{\hermq}{\mN} \times \hermv \right) \)										& \( \Op{ 12 } = \hermS{\chi} \cdot \left( \hermS{\N} \times \hermv \right) \)																				\\
			\( \Op{ 6 } = \left( \hermS{\chi} \cdot \frac{\hermq}{\mN} \right) \left( \hermS{\N} \cdot \frac{\hermq}{\mN} \right) \)	& \( \Op{ 13 } = i \left( \hermS{\chi} \cdot \hermv \right) \left( \hermS{\N} \cdot \frac{\hermq}{\mN} \right) \)											\\
			\( \Op{ 7 } = \hermS{\N} \cdot \hermv \)																					& \( \Op{ 14 } = i \left( \hermS{\chi} \cdot \frac{\hermq}{\mN} \right) \left( \hermS{\N} \cdot \hermv \right) \)											\\
			\( \Op{ 8 } = \hermS{\chi} \cdot \hermv \)																					& \( \Op{ 15 } = - \left( \hermS{\chi} \cdot \frac{\hermq}{\mN} \right) \left[ \left( \hermS{\N} \times \hermv \right) \cdot \frac{\hermq}{\mN} \right] \)	\\
			\bottomrule
		\end{tabular}
		\caption[Fourteen linearly independent \acp{NREO}]{All fourteen linearly independent \acp{NREO}, adopted from~\cite{Catena:2015uha}.}%
		\label{tab:operators}
	\end{table*}

	The interaction operators that can be constructed from the basis~\eqref{eqn:HermitianOps} are shown in Tab.~\ref{tab:operators}.
	Further, the second operator \( \Op{2} = {\left(\hermv\right)}^2 \) will be omitted in following with~\cite{Anand:2013yka} as it cannot be leading order in a non-relativistic limit of a relativistic model of \ac{DM} interactions.
	As we are interested in \ac{DM} in the galactic halo, the \ac{DM} has velocities below the escape velocity (\(v_\text{esc} = \qty{544}{\km \per \s}\)~\cite{Frandsen:2011gi}), so such non-relativistic limits are safe to consider. 
	It is worth noting that the first (\( \Op{1} \)) and fourth (\( \Op{4} \)) operators respectively correspond to the standard, constant \ac{SI} and \ac{SD} interactions.

	Generically, the low-energy limit of an effective DM-SM interaction can be parameterised as a linear combination of the operators in Tab.~\ref{tab:operators}:
	\begin{equation}
		\mathbf{\widehat{\mathcal{H}}}(\mathbf{r}) = \sum_{\tau=0,1} \sum_{k=1}^{15} \couple{k} \Op{k}(\mathbf{r}) \, t^{\tau} \, ,
		\label{eqn:genHamIso}
	\end{equation}		
	where \( t^{\tau} \) represents the coupling to weak isospin: \( t^0 = \mathbbm{1} \) represents isoscalar couplings, and \( t^1 = \tau_3 \) (the third Pauli matrix) for isovector couplings.
	The couplings can be projected onto the basis of protons or neutrons: \(c_k^{p/n} = (c_k^0 \pm c_k^1)/2 \).

	Each DM-nucleon interaction operator couples to a different combination of states in the basis of the \textit{nucleus}.
	For example \( \Op{3} \) contains the nucleon spin \( \hermS{N} \), but does not couple to the total nucleus spin, but to the spin-orbit interaction.
	This yields nonzero effective couplings between this ``spin-dependent'' particle and e.g.\ the spin-zero \( ^{40}\text{Ar} \) nucleus.

	As an example of how such operators could arise, we may look at \( \Op{10} \), which could arise from a parity-violating \ac{DM}-quark interaction mediated by a heavy scalar~\cite{Dent:2015zpa}:
	\begin{multline}
		\mathcal{L} \supset \lambda_1 \phi \bar{\chi}\chi - i h_2 \phi \bar{q} \gamma^5 q \rightarrow \\
			\widehat{\mathcal{H}} \supset \left( \couple[0]{10} t^0 + \couple[1]{10} t^1 \right) \Op{10} ,
		\label{eqn:O10_L}
	\end{multline}
	in this case \( \Op{10} \) is the leading operator contributing to this interaction.
	Here \( \lambda_1 \) and \( h_2 \) are coupling parameters in the EFT Lagrangian, \( \chi \) is the \ac{DM}, \( q \) are quarks, \( \phi \) is a scalar mediator between the \ac{DM} and quarks.
	
	The \ac{DM}-nucleus scattering cross section can be factored into a DM response function \(R^{\tau \tau'}\), which depends on the combination of couplings, and a nuclear response function \(W^{\tau \tau'}\).

	The \ac{DM} response functions take a general form:
	\begin{align}
		\begin{split}
			R^{\tau \tau^\prime} & \left(\vtangsq, \frac{\hermq^{2}}{\mN^2}, \left\{ \couple{i} \couple[\tau^\prime]{j} \right\} \right) = \\ 
			& \sum_{n=0}^1 \sum_{m=0}^2 A_{n,m} \left( \vtangsq \right)^n \left( \frac{\hermq^{2}}{\mN^2} \right)^m \left\{ \couple{i} \couple[\tau^\prime]{j} \right\}_{n,m} ,
		\end{split}
		\label{eqn:generalDarkResponse}
	\end{align}
	where they generically appear as quadratic combinations of the velocity (indexed by \( n \)) and transferred momentum (indexed by \( m \)) with prefactors \( A_{n,m} \) and pairs of coupling constants.
	The coupling constant pairs are bounded by curly braces, where each pair is associated with a unique combination of velocity and momentum powers (\( n, m \)), and each pair may (\( i = j \)) or may not (\( i \neq j \)) correspond to the same operator index.
	The specific form for all \ac{DM} response functions can be found in Appendix~A of~\cite{Catena:2015uha}.
	The nuclear response functions that we use were computed using nuclear shell models and tabulated in Ref.~\cite{Catena:2015uha}.
	These are fitted to the general form:
	\begin{equation}
		W^{\tau \tau^\prime}(y) = \exp(-2Dy) \sum_{i=0}^6 A^{\tau \tau^\prime}_i y^i ,
		\label{eqn:generalNuclearResponse}
	\end{equation}
	where the exponential may (\( D=1 \), not hydrogen) or may not (\( D=0 \), hydrogen) contribute to a given function depending on which isotope is being calculated, and the coefficients \( A^{\tau \tau^\prime}_i \) corresponding to each power of \( y \) and each isospin index (\( \tau, \tau^\prime \)) must be computed separately for each isotope.

	The parameter \( y \) is defined as:
	\begin{equation}
		y = {\left( \frac{b q}{2} \right)}^2 ,
		\label{eqn:yDef}
	\end{equation}
	and \( b \) is a length parameter corresponding to the scale of the nucleus:
	\begin{equation}
		b = \sqrt{\frac{41.467}{45A^{-\frac{1}{3}} - 25A^{-\frac{2}{3}}}} \text{ fm} ,
		\label{eqn:bDef}
	\end{equation}
	where \( A \) is the atomic mass number of the isotope.

	While the couplings \( \couple{i} \) are in principle derived from a UV-complete theory, they are difficult to compare directly with other \ac{DM} search results.
	We may define an effective cross section (as in e.g.~\cite{DEAP:2020iwi}):
	\begin{equation}
		\sigma_N = \frac{{\left( \couple[0]{i} \mu_{p} \right)}^2}{4\pi} ,
		\label{eqn:effXSec}
	\end{equation}
	where \( \mu_{p} \) is the proton-\ac{DM} reduced mass, and we consider an isoscalar (\(\couple[0]{i} \neq 0 \,,\, \couple[1]{i} = 0\)) interaction, making use of \( \couple[p]{i} = \left(\couple[0]{i} + \couple[1]{i}\right) / 2 \). While isospin violation has been invoked in many contexts (mainly to remove constraints from specific experiments), we limit ourselves to isospin-conserving interactions, as it is a minimal assumption on how the dark sector couples to quarks. On the practical side, it means one fewer free parameter per operator. 
	In the case of a constant cross section these directly correspond to the constant \ac{SI} (\( \Op{1} \)) and \ac{SD} (\( \Op{4} \)) interactions, though to match with the typical definition of \ac{SD} the effective cross section expression in~\eqref{eqn:effXSec} is multiplied by \(3/16\)~\cite{Anand:2013yka,ParticleDataGroup:2022pth}.
	It is worth emphasizing that Eq.~\eqref{eqn:effXSec} loses its immediate significance for velocity or momentum-dependent interactions.

\section{Stellar Capture and Annihilation}\label{Sec:Capture}

	Below, we describe the process of gravitational capture of halo \ac{DM} by the Sun.
	We take the \ac{DM} to follow a Maxwell-Boltzmann distribution.
	In the frame of the Sun~\cite{Scott:2008ns}:
	\begin{multline}
		f_\odot(u) = {\left(\frac{3}{2}\right)}^{3/2}\frac{4\rho_\chi u^2}{\pi^{1/2}m_\chi u_0^3} \\
			\exp\left(-\frac{3 (u_\odot^2+u^2)}{2u_0^2} \right) \frac{\sinh (3 u u_\odot/u_0^2)}{3 u u_\odot /u_0^2} ,
		\label{eqn:maxBolt}
	\end{multline}
	where \( \rho_\chi \) and \( m_\chi \) are the \ac{DM} density and mass, \( u \) is the \ac{DM} velocity, \( u_0 \) is the velocity dispersion, and \( u_\odot \) is the Sun's velocity.

	The particle can be scattered elastically to a velocity below the local escape velocity, allowing it to be gravitationally bound to the Sun.
	The capture rate can be written as an integral over the Sun's radius \( R_{\odot} \) and the second over the \ac{DM} velocity distribution~\cite{Gould:1987ir}:
	\begin{equation}
		C = 4\pi \int_{0}^{R_{\odot}} r^2 \diff r \int_{0}^{\infty} \frac{f(u)}{u} w(r) \Omega_{v}^{-}(w(r)) \diff u ,
		\label{eqn:capRate}
	\end{equation}
	where \( \Omega_{v}^{-}(w(r)) \) is the rate at which \ac{DM} is scattered from speed \(w(r) = \sqrt{u^2 + {v_{esc}(r)}^2}\) below the local escape velocity \(v_{esc}(r)\):
	\begin{multline}
		\Omega_{v}^{-}(w) = \sum_i n_i w \Theta\left( \frac{\mu_i}{\mu^2_{+,i}} - \frac{u^2}{w^2} \right) \\
			\int_{E_k u^2/w^2}^{E_k \mu_i/\mu_{+,i}^2} \diff E_R \frac{\diff \sigma_{i}}{\diff E_R}\left(w^2,q^2\right) .
		\label{eqn:Omega}
	\end{multline}
	The sum is over the different isotopes found in the Sun, where \( n_i \) is the \( i^{\text{th}}\) isotope's abundance as a function of the radius, \( E_k = m_\chi w^2 / 2 \) is the kinetic energy of the \ac{DM} particle, and
	\begin{equation}
		\mu_i\equiv \frac{m_\chi}{m_i},\, \qquad\qquad \mu_{\pm,i}\equiv \frac{\mu_i\pm1}{2}\,.
		\label{eqn:muDef}
	\end{equation}

	\( \Omega_{v}^{-} \) depends on the differential cross section \( {\diff \sigma}/{\diff E_R} \).

	The integrals in Eqs.~\ref{eqn:capRate}-\ref{eqn:Omega} can be factored into terms of the form \(\text{(polynomial in }y\text{)} \, e^{-y}\) which have solutions proportional to gamma functions and are detailed in Ref.~\cite{Vincent:2015gqa}.
	By performing this factorisation, and gathering terms by order in \(y\), where \( y \propto q^2 \) is defined in Eq.~\ref{eqn:yDef}, the integral~\eqref{eqn:Omega} can be sped up significantly.

	The geometric limit \(\sigma_\text{max} = \pi R_\odot^2(t)\) for which the star captures all \ac{DM} it encounters leads to a maximum capture rate~\cite{Vincent:2015gqa}:
	\begin{equation}
		\begin{split}
			C_\text{max}(t)	& = \pi R_\odot^2(t) \int_0^\infty \frac{f_\odot(u)}{u}w^2(u,R_\odot) \mathrm{d} u \label{eqn:capcutoff} \\
							& = \frac{1}{3}\pi\frac{\rho_\chi}{m_\chi}R_\odot^2(t)\bigg(e^{-\frac{3}{2}\frac{u_\odot^2}{u_0^2}}\sqrt\frac{6}{\pi}u_0 \\
							& + \frac{6G_\mathrm{N}M_\odot + R_\odot(u_0^2 + 3 u_\odot^2)}{R_\odot u_\odot}\mathrm{Erf}{\left[\sqrt{\frac{3}{2}}\frac{u_\odot}{u_0}\right]} \bigg) \, ,
		\end{split}
	\end{equation}
	Where \( G_\mathrm{N} \) is Newton's gravitational constant.
	Eq.~\eqref{eqn:capcutoff} includes gravitational focusing.
	The overall capture rate is then the minimum of Eq.~\eqref{eqn:capRate} and Eq.~\eqref{eqn:capcutoff}.

	\subsection{Dark Matter in the Sun}\label{sub:DMinSun}

		For \ac{DM} masses above approximately \(m_\chi \simeq 4 \) GeV, evaporation rapidly becomes negligible~\cite{Busoni:2017mhe}.
		As here our most relevant constraints from annihilating \ac{DM} only exist at \( \mathcal{O}(\qty{100}{\GeV}) \), this evaporation effect can be safely ignored.
 		In addition, for the cross sections considered here, we have checked that the capture and annihilation rates in the Sun will have equilibrated by now.
		The annihilation rate is therefore equal to the capture rate, and thus proportional to the \ac{DM}-nucleon scattering cross section.
		Annihilation to any \ac{SM} product will eventually yield a neutrino flux.
		We will examine a few representative channels: \( \chi \chi \rightarrow \bb \), \( \WW \) and \( \tautau \).
		Once neutrinos are produced, they will quickly escape the core of the Sun and can be detected at Earth-based observatories including \ic[], \antares[], \deepcore[], and \superk[]~\cite{IceCube:2012ugg,IceCube:2012fvn,IceCube:2016yoy,ANTARES:2016xuh,IceCube:2021xzo,Super-Kamiokande:1998qwk}.

\section{Methods}\label{sec:Methods}

	\subsection{GAMBIT}

		Our global scan uses \ac{GAMBIT}~\cite{GAMBIT:2017yxo}, a global fitting package that allows users to specify models, perform likelihood calculations based on physical observables, and scan large parameter spaces with state-of-the-art scanners.
		GAMBIT has the ability to compute likelihoods directly, but also to ``backend'' existing public code such as DarkSUSY~\cite{Gondolo:2004sc,Bringmann:2018lay}, DDCalc~\cite{GAMBITDarkMatterWorkgroup:2017fax,GAMBIT:2018eea} and Capt'n General~\cite{Kozar:2021iur}.
		GAMBIT is publicly available at \url{https://gambitbsm.org}.
		For this work, we used a modified version of 2.3.1.

		\ac{GAMBIT} is made of several modules, each corresponding to a class of observables like DecayBit~\cite{GAMBITModelsWorkgroup:2017ilg} and DarkBit~\cite{GAMBITDarkMatterWorkgroup:2017fax} which handle decay rates and \ac{DM} physics respectively, or computations like the ScannerBit module~\cite{Martinez:2017lzg}, which is designed to coordinate the exploration of the likelihood parameter space through the use of \ac{MCMC} (GReAT~\cite{Putze:2014aba} and T-Walk~\cite{Martinez:2017lzg}), nested sampling (MultiNest~\cite{Feroz:2008xx}), or \ac{DE} (Diver~\cite{Martinez:2017lzg}, which we use for this work) packages.
		\ac{GAMBIT} is designed modularly such that observables can be used in arbitrary functions defined using genericized dependencies and output capabilities.
		The functions of \ac{GAMBIT} can request an arbitrary number and variety of dependencies.
		Each function then has an output capability, which further functions can use as input dependencies.
		In addition, \ac{GAMBIT} can also make use of external code packages called backends for capability calculation.
		Capt'n General is one of these backends.

		This structure allows \ac{GAMBIT} to determine a call order of functions to calculate likelihoods from desired observables using as many intermediate steps as required.
		This dependency resolver constructs an acyclic graph to organize the function calls such that the capabilities are properly passed down the resolved tree to calculate likelihoods.
		In addition, any ambiguous call order is resolved intelligently by making faster function calls first such that points can be invalidated by quicker likelihood calculations before slower ones.
		With a call order established, ScannerBit controls the exploration of the parameter space according to the selected scanning algorithm.
		Our scans will use DarkBit, as well as SpecBit and DecayBit to read in \ac{SM} inputs and generate decay rates for the catalogue of annihilation products.

		For this work, the GAMBIT front end required additions to accommodate the \ac{NREO} parametrization.
		This included incorporating the couplings to the inbuilt DDCalc coupling dependencies, and creating a catalogue of particles that can be produced in \ac{DM} annihilation.

	\subsection{Direct Detection Likelihoods}

		The \ac{DD} experiments are handled by \dac{GAMBIT} through the use of the DarkBit module~\cite{GAMBITDarkMatterWorkgroup:2017fax}, which handles the calculation of event rates in each experiment from the input couplings through the use of DDCalc~\cite{GAMBITDarkMatterWorkgroup:2017fax,GAMBIT:2018eea}.

		DDCalc is a backend for recasting direct search limits.
		It calculates the predicted rates and likelihoods for a variety of experiments, given an input dark matter and halo model.
		Data is compared with these predictions using a binned Poisson likelihood.
		DDCalc models experimental backgrounds when made available by each collaboration; otherwise it computes a one-sided likelihood that can be used for exclusion (disfavouring models that predict larger event numbers than observed), but cannot identify a preferred signal.
		As of DDCalc v2.0~\cite{GAMBIT:2018eea}, the code allows for the use of NREOs directly, and incorporates the nuclear response functions computed in~\cite{Anand:2013yka} necessary to evaluate Eq.~\eqref{eqn:generalNuclearResponse} in the context of DD experiments.
		As stated earlier, we use the likelihoods for \allDD[c].
		Details of how the results of each experimental analysis is recast can be found in Appendix A of Ref.~\cite{GAMBIT:2018eea}.

	\subsection{Solar Capture}

		We compute the solar capture rate using the Capt'n General code~\cite{Kozar:2021iur}.
		Capt'n General is written in Fortran 90.
		It makes use of a set of \ac{DM} halo parameters, a solar model with elemental abundances, and a set of \ac{DM} parameters, to calculate the \ac{DM} capture rate in the Sun.
		It is written to interface as an external library in software such as GAMBIT or MESA.\@
		The first version of Capt'n General parametrised non-constant \ac{DM}-nucleus interactions using the formalism of~\cite{Vincent:2015gqa}, where the \ac{DM}-nucleus interactions were scaled with momentum and velocity, i.e.\ \( \sigma = \sigma_0 {(v/v_0)}^{2n} \) or \( \sigma = \sigma_0 {(q/q_0)}^{2n} \), where \( v_0 \) and \( q_0 \) are arbitrary reference values. 
		We extended Capt'n General to make use of the \ac{NREO} formalism as discussed in Sec.~\ref{sec:Theory}~\cite{Kozar:2021iur}.\footnote{\href{https://github.com/aaronvincent/captngen}{github.com/aaronvincent/captngen}}

		A key part of the software uses \acp{GFFI} to calculate integrals of the form\footnote{The actual implementation is based on the variable \( E_R \) rather than \( y \), so the exact form of the GFFI in the code more closely follows Ref.~\cite{Vincent:2015gqa}. Because of this, the GFFI results presented here each differ by a constant factor.} 
		\begin{equation}
			GFFI = \int_{y_\text{min}}^{y_\text{max}} {y}^n e^{-D 2 y} \diff y.
			\label{eqn:GFFI}
		\end{equation}
		as in Eqn.~\eqref{eqn:Omega}, with \( y_\text{min} = b^2u^2m_{i} m_\chi/4 \) and \( y_\text{max} ={b^2 w^2 m_\chi^2}/{4 \mu_{+,i}^2} \).
		For each operator, the dark matter response function and the nuclear form factors come with sums over powers of \( y \).
		Capt'n General first groups integrals into powers of \( y \), evaluates the results of Eq.~\eqref{eqn:GFFI}, and performs the sums in Eqs.~\eqref{eqn:generalDarkResponse} and~\eqref{eqn:generalNuclearResponse}.

		When the target is hydrogen (\( D = 1 \)), 
		\begin{equation}
			\begin{split}
				&GFFI_{n\ne0,\mathrm{H}} = \\
				&
				\begin{cases}
					\frac{1}{1+n} {\left(\frac{b^2 w^2 m_i m_\chi}{4}\right)}^{n+1} \left[{\left(\frac{\mu}{\mu_{+}^2}\right)}^{n+1} - {\left(\frac{u^2}{w^2}\right)}^{n+1}\right]	&(n \neq -1) \\
					\ln \left(\frac{\mu}{\mu_+^2}\frac{w^2}{u^2}\right)													&(n = -1).
				\end{cases}
				,
			\end{split}
			\label{eqn:GFFI_H}
		\end{equation}
		When \( D = 1 \),
		\begin{equation}
			\begin{split}
				&GFFI_{n\ne0,i\ne\mathrm{H}} = \\
				&\!\frac{1}{2^{n+1}}\left[ \Gamma\!\left(\! 1+n,\frac{b^2 u^2 m_i m_\chi}{2} \!\right) - \Gamma\!\left(\! 1+n,\frac{b^2 w^2 m_\chi^2}{2 \mu_{+,i}^2} \!\right) \right], 
			\end{split}
			\label{eqn:GFFI_A}
		\end{equation}
		where \( \Gamma(m,x) \) is the (upper) incomplete gamma function.

		Once integrals over \( y \propto q^2 \) have been performed, integrals over velocity are done numerically using a QUADPACK~\cite{piessens1983,Gondolo:2004sc}, after which the integral over the volume of the star is performed with trapezoidal integration.
		We precompute a set of prefactors that each are a sum of all calculated constants from their respective \ac{GFFI} velocity and momentum combinations.
		This set of prefactors can then be used to make the minimum number of calls to the numerical integration routine to complete the velocity (\( u \)) integral of Eqn.~\eqref{eqn:capRate}, allowing the code to entirely skip some calls depending on which prefactors end up being zero.

		Capt'n General is implemented as a backend to \dac{GAMBIT}, and is called once per likelihood evaluation.The current release of \dac{GAMBIT} 2.3.1 includes event-level neutrino likelihoods from the 3-year 79-string IceCube (IC79) data~\cite{IceCube:2012ugg,IceCube:2016yoy} via nulike~\cite{IceCube:2012fvn}, with neutrino spectra generated and propagated to Earth by DarkSUSY~\cite{Gondolo:2004sc,Bringmann:2018lay}.
		We supplement these with the most recently-presented IceCube analysis using 8 years of data and the full 86-string detector~\cite{Meighen-Berger:2022} at masses above \qty{300}{\GeV}, and use the older IC79 dataset for masses below.
		These were presented as 90\% limits on the spin-independent DM-nucleon scattering cross section as a function of DM mass.
		To incorporate the~\cite{Meighen-Berger:2022} results, we map their constraints on a constant, spin-dependent dark matter-nucleon cross section onto a constraint on the capture rate using Capt'n General.
		This allows us to set limits for any arbitrary combination of NREO operators and couplings.
		In the same way, we also include recently-presented \antares[c] limits, and the most recent \deepcore[c] and \superk[c] limits.

	\subsection{Differential Evolution Scanner}

		We employ the Diver~\cite{Martinez:2017lzg} scanner, a \ac{DE} algorithm for parameter space sampling.
		Ref.~\cite{Martinez:2017lzg,DarkMachinesHighDimensionalSamplingGroup:2021wkt} showed that Diver is much more efficient than a range of other algorithms at finding features in parameter space, with the drawback of not yielding a posterior likelihood distribution.
		This will not be an issue, as we will present results in terms of frequentist profile likelihoods.

		The \ac{DE} algorithm uses a set of points placed in the parameter space which are used to map the likelihood.
		This process occurs in three steps: mutation, crossover, and selection.

		\ac{DE} starts from an initial random population of points.
		The first step is mutation: for each point three other points are selected randomly, two of which are used to create a donor vector and applied to the third point with a scaling factor.
		In the second step (crossover) the donor vector is mixed with the original point's vector to produce a trial vector.
		This is done by stepping through each component of the trial vector and randomly choosing either the original's or the donor's component.
		After each component is chosen, if this trial vector mixture is identical to the original point then a random component is chosen to be switched for the donor's.
		In the third step (selection) the original point's and trial vector's likelihood are compared, and the one with the better value is retained for the next set of initial points -- if the likelihood is equal, the trial vector is chosen to allow the algorithm to explore flat regions~\cite{Martinez:2017lzg}. 

		\ac{DE} is useful here as it is focused on mapping the contours of the parameter space being scanned, rather than identifying global minima.
		As Diver only accepts updated points that improve the population's likelihood, the algorithm tends to converge faster than alternatives that can accept inferior points based on random chance such as \acp{MCMC}.\@

		Varying all NREO operators simultaneously would not yield much constraining power.
		We separate the \dac{GAMBIT} scan into fourteen sets of parameter space scans of each isoscalar \ac{NREO} in Tab.~\ref{tab:operators}, where each isoscalar operator had three scan versions depending on annihilation channel: one with annihilation to \( \bb \) which yields a soft neutrino spectrum, a second with annihilation to \( \WW \) (harder), and a third with annihilation to \( \tautau \) (hardest).
		Annihilation to other final states, or a combination of these, would yield constraints intermediate between these cases, except for the direct annihilation to neutrinos.\footnote{Indeed, direct annihilation to $\bar \nu \nu$ would yield the strongest constraints.  However, models that predict a strong enough coupling to quarks (necessary for capture), while retaining a large annihilation rate directly to neutrinos, would necessitate some tuning. Even in the case of e.g. Kaluza Klein dark matter which yields such a channel, the dominant neutrino flux comes from charged $\tau$ decay \cite{Hooper:2002gs}.} Results for these final states are  available in  \ic[c] analyses used here.
		Including isovector couplings, equivalent to varying the ratio of proton-to-neutron couplings, would change the relative constraining power of direct detection experiments examined here (to see this interplay, see e.g.~\cite{Brenner:2022qku, Cheek:2023zhv}).
		To avoid the additional computational time associated with this extra parameter in each model, we restrict ourselves to isoscalar models (equal coupling to protons and neutrons).
		We fix the annihilation cross-section to \qty{3e-26}{\cm^3 \, \s^{-1}}.
		In each of the forty-two individual scans the \ac{DM} mass was allowed to vary from \qty{1}{\GeV} to \qty{10}{\TeV} with a log-flat prior.
		In addition, we chose each scan's coupling range such that the scans captured the region in which the 90\% CL resided, with the beginning and end points chosen by first referencing a coarser and quicker initial scan.
		We show these coupling parameter ranges in the first section of Tab.~\ref{tab:parameters}.
		We preformed each of these scans using a population of \num{e4}, and a convergence threshold of \num{e-3} in Diver~\cite{Martinez:2017lzg}.

		In addition, each scan also included three nuisance parameters quantifying the \ac{DM} velocity distribution: the halo velocity dispersion \( v_0 \), the galactic escape velocity \( v_\text{esc} \) and the Sun's galactic rotational velocity \( v_\text{rot} \).
		We set the local dark matter density to \qty{0.5}{\GeV{} \, \cm^{-3}}, and did not vary it, as it is completely degenerate with the (square of the) couplings.
		We allowed each nuisance parameter to vary with a gaussian prior with central values and widths informed by Refs.~\cite{Reid:2014boa} and~\cite{Deason:2019tpl}, as used in Ref.~\cite{GAMBIT:2021rlp}.
		These are shown in the second section of Tab.~\ref{tab:parameters}.
		We processed the resulting Diver scans using Pippi~\cite{Scott:2012qh} and plotted the results using Matplotlib~\cite{Hunter:2007ouj} and adjustText~\cite{flyamer:2023}.

		\begin{table}[htb]
			\centering
			\begin{tabular}{cc}
				\midrule
				\multicolumn{2}{c}{Coupling parameters (\unit{\GeV^{-2}})}					\\
				\midrule
				\( \log_{10}(c_{1}^{0})   \)	& (\numrange[range-phrase={, }]{-10}{-6 })	\\
				\( \log_{10}(c_{3}^{0})   \)	& (\numrange[range-phrase={, }]{-6 }{-3 })	\\
				\( \log_{10}(c_{4}^{0})   \)	& (\numrange[range-phrase={, }]{-8 }{-3 })	\\
				\( \log_{10}(c_{5}^{0})   \)	& (\numrange[range-phrase={, }]{-5 }{-2 })	\\
				\( \log_{10}(c_{6}^{0})   \)	& (\numrange[range-phrase={, }]{-5 }{-1 })	\\
				\( \log_{10}(c_{7}^{0})   \)	& (\numrange[range-phrase={, }]{-4 }{-1 })	\\
				\( \log_{10}(c_{8}^{0})   \)	& (\numrange[range-phrase={, }]{-6 }{-4 })	\\
				\( \log_{10}(c_{9}^{0})   \)	& (\numrange[range-phrase={, }]{-6 }{-1 })	\\
				\( \log_{10}(c_{10}^{0})  \)	& (\numrange[range-phrase={, }]{-6 }{-2 })	\\
				\( \log_{10}(c_{11}^{0})  \)	& (\numrange[range-phrase={, }]{-9 }{-5 })	\\
				\( \log_{10}(c_{12}^{0})  \)	& (\numrange[range-phrase={, }]{-8 }{-4 })	\\
				\( \log_{10}(c_{13}^{0})  \)	& (\numrange[range-phrase={, }]{-5 }{-1 })	\\
				\( \log_{10}(c_{14}^{0})  \)	& (\numrange[range-phrase={, }]{-3 }{+1 })	\\
				\( \log_{10}(c_{15}^{0})  \)	& (\numrange[range-phrase={, }]{-5 }{-2 })	\\
				\\
				\midrule
				\multicolumn{2}{c}{Common model parameters}				\\
				\midrule
				\( \log_{10}(m_\text{dm}) \) (\unit{\GeV})	& (\numrange[range-phrase={, }]{+0 }{+4 })	\\
				\( \rho_0 \) (\unit{\GeV{} \, \cm^{-3}})	& \num{0.5}				\\							
				\( v_0 \) (\unit{\km{} \, \sec^{-1}})		& (\numrange[range-phrase={, }]{216}{264})	\\
				\( v_{rot} \) (\unit{\km{} \, \sec^{-1}})	& (\numrange[range-phrase={, }]{216}{264})	\\
				\( v_{esc} \) (\unit{\km{} \, \sec^{-1}})	& (\numrange[range-phrase={, }]{453}{603})
			\end{tabular}
			\caption[]{Table of parameters used in the global Diver scans.
			Each individual scan is over a single coupling, while every scan includes the common parameters listed in the lower portion of the table.
			The dark matter mass (\( m_\text{dm} \)) and couplings (\( c_{i}^{0} \)) have log-flat priors, and the halo parameters (\( v_0 \), \( v_{rot} \), and \( v_{esc} \)) have Gaussian priors, where the ranges indicate the \( 3\text{--}\sigma \) ranges of these values.
			The halo parameters have the same range as used in~\cite{GAMBIT:2021rlp}, where the ranges of \( v_0 \), \( v_{rot} \) and \( v_{esc} \) are from~\cite{Reid:2014boa} and~\cite{Deason:2019tpl} respectively.}%
			\label{tab:parameters}
		\end{table}

\section{Results}\label{sec:Results}

\begin{figure*}[htb]
		\centering
		\centerline{\includegraphics[width=\paperwidth]{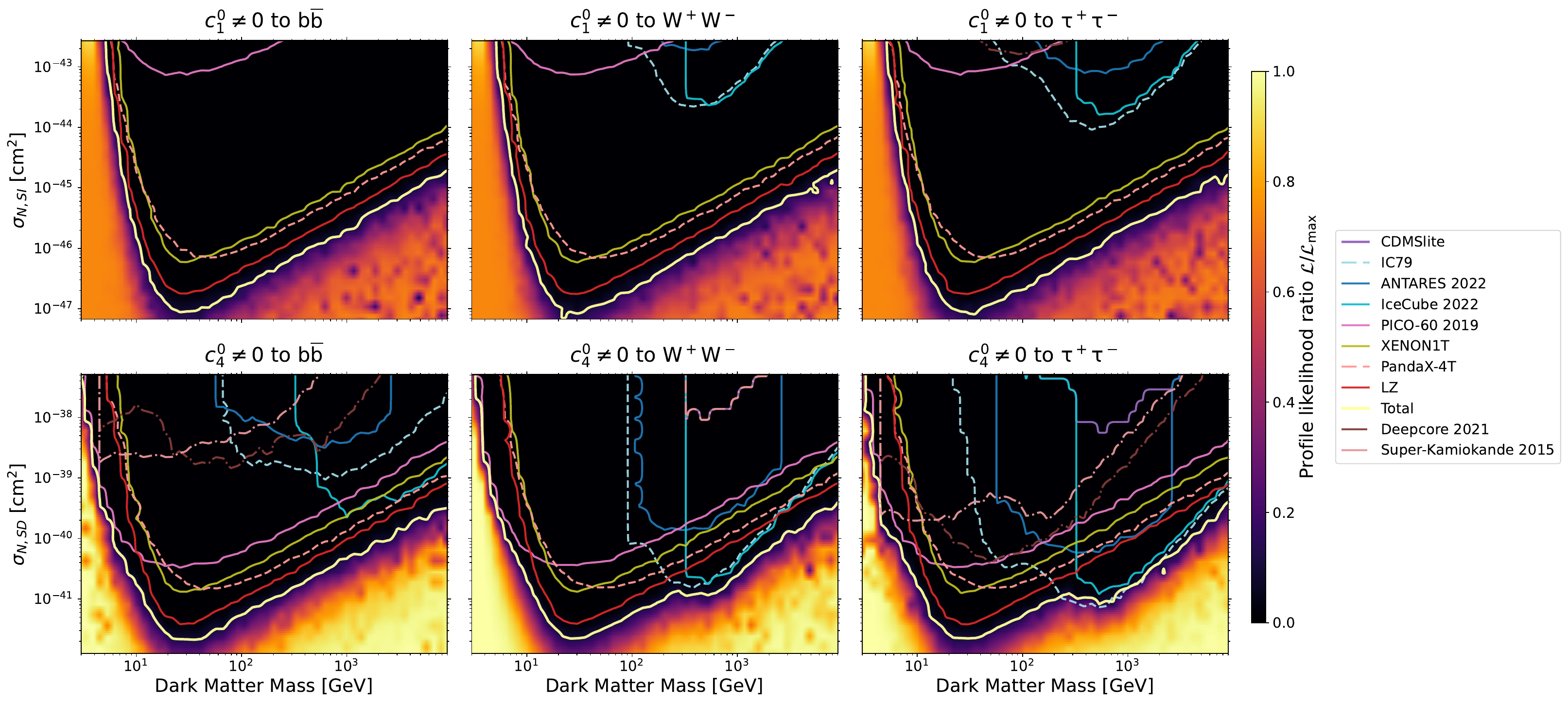}}
		\caption{Profile likelihood ratios in the plane of dark matter mass and effective cross section for \( \couple[0]{1} \) (top) and \( \couple[0]{4} \) (bottom).
			These respectively correspond to constant \ac{SI} and \ac{SD} cross sections (Eq.~\ref{eqn:effXSec}).
			Left: annihilation channel to \( \bb \), center: \( \WW \), and right: \( \tautau \).
			The lines indicate the 90\% CL contours of the total and each experiment's contribution to the total.
			The coloured histogram represents the profile likelihood ratio \( \mathcal{L} / \mathcal{L}_\text{max} \) of the total.
			We chose to only display a selection of experiments here that contribute most to the total to reduce visual noise, with complete plots shown in Appendix~\ref{appendix:allscans}.
			The IC79 line \emph{does not} contribute to the total above \qty{300}{\GeV} as it has overlap with IceCube (2022).
			The contribution of solar neutrinos becomes more dominant when \ac{DD} experiments don't receive the \(A^2\) enhancement and a harder annihilation channel is chosen.
			These effects manifest in the central and right-hand \ac{SD} plots as the blue \ic[c] curves dropping below the red \lz[c] curve in the \(\qty{500}{\GeV} \text{ to } \qty{1}{\TeV}\) region.}%
		\label{fig:c1c4Compare}
	\end{figure*}

\begin{figure*}[htb]
		\centering
		\centerline{\includegraphics[width=\paperwidth]{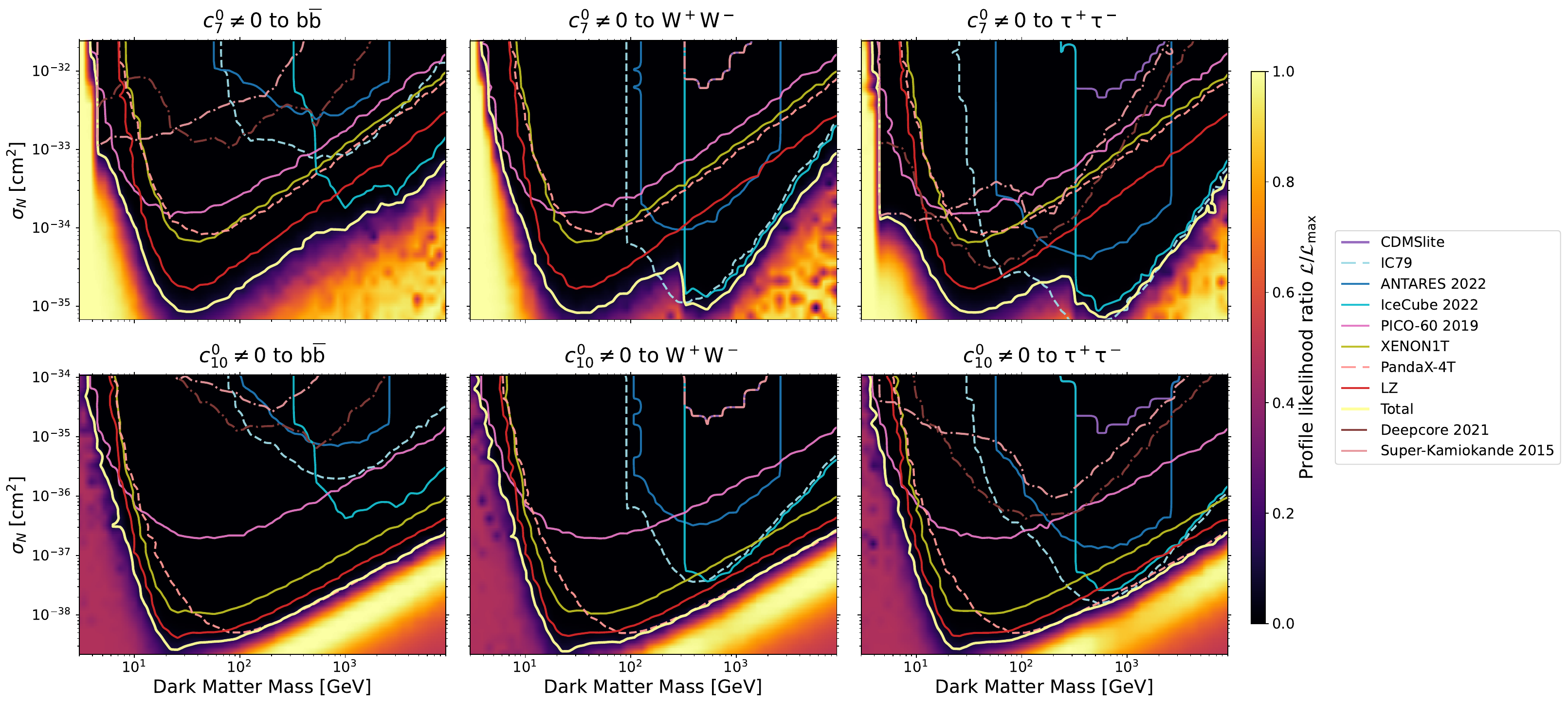}}
		\caption[Profile likelihood of \({\couple[0]{7}}\) and \({\couple[0]{10}}\)]{Same as Fig.~\ref{fig:c1c4Compare}, but for \({\couple[0]{7}}\) and \({\couple[0]{10}}\).
		The slightly better fit just below the exclusion line in the bottom panel is likely due to an underfluctuation in the LZ event rate at low recoil energies; see main text for details.}%
		\label{fig:c7c10Compare}
	\end{figure*}

\begin{figure*}[htb]
		\centering
		\centerline{\includegraphics[width=\paperwidth]{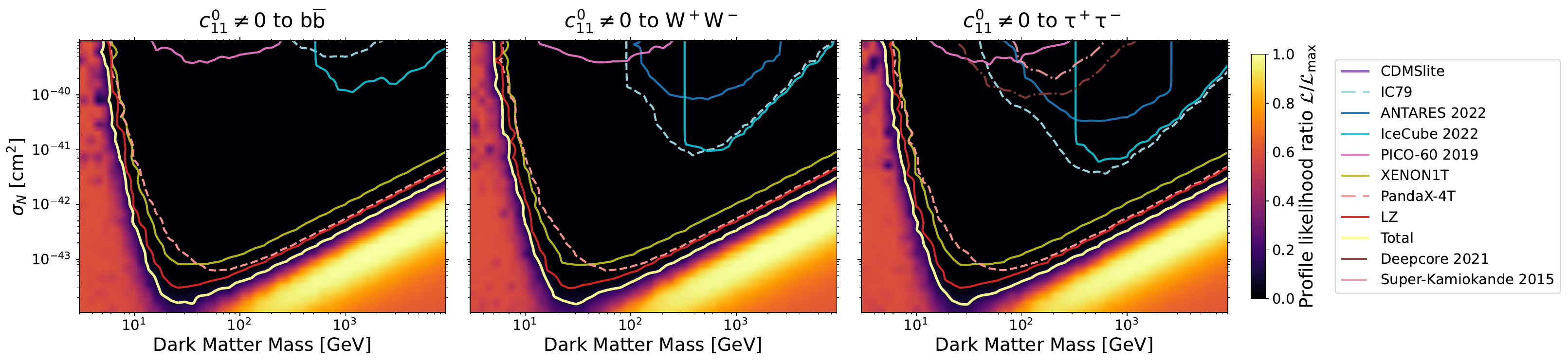}}
		\caption[Profile likelihood of \({\couple[0]{11}}\)]{Same as Fig.~\ref{fig:c1c4Compare}, but for \({\couple[0]{11}}\).}%
		\label{fig:double11best}
	\end{figure*}

	The results are presented in Figs.~\ref{fig:c1c4Compare},~\ref{fig:c7c10Compare}, and~\ref{fig:double11best} as profiled likelihoods, where we show 90\% CL contours, and the black (low-likelihood) regions are excluded.
	We present three panels for each of the isoscalar couplings corresponding to the \( \bb \), \( \WW \), and \( \tautau \) annihilation channels.
	We show the 90\% CL contour contributed by individual experiments to the total likelihood, which also has its own 90\% CL contour shown.
	For clarity, we only show the constituent 90\% CL contours for experiments that dominate in contribution to the total likelihood.

 	We show neutrino telescope constraints from IC79, ANTARES, IceCube (2022), \deepcore[], and \superk[].
	Note that the IC79 and IceCube (2022) samples have overlap and are thus not independent.
	However, the IceCube (2022) analysis \emph{does not} extend below \qty{300}{\GeV}.
	Therefore, we include the IC79 results in the total log-likelihood only below \qty{300}{\GeV}, and the IceCube (2022) results only above \qty{300}{\GeV}.

	In the main text, we will present constraints on \( \couple[0]{1} \) and \( \couple[0]{4} \), which respectively yield the \ac{SI} and \ac{SD} constant cross sections, in addition to \( \couple[0]{7} \), \( \couple[0]{10} \) and \( \couple[0]{11} \), which illustrate the different ways in which \ac{DD} and neutrino telescopes complement each other.
	The full results for the 14 operators shown in Tab.~\ref{tab:operators} are presented in Appendix~\ref{appendix:allscans}, including individual likelihood contributions of \textit{every} experiment.

	The results of the scans for \( \couple[0]{1} \) and \( \couple[0]{4} \) are shown in Fig.~\ref{fig:c1c4Compare}, plotted using both the coupling and the effective cross section as expressed in Eq.~\eqref{eqn:effXSec}.

	Not unexpectedly, direct detection experiments dominate spin-independent limits due to the coherent \( A^2 \) enhancement of the cross section, whereas neutrino telescopes remain competitive in spin-dependent searches for some channels at higher DM masses.
	Harder spectra (\( \WW \) and \( \tautau \)) yield more competitive results.

	The likelihoods are fairly flat outside the excluded region, indicating the absence of a preferred region of parameter space.
	
	Some of the strongest constraints from neutrino telescopes are on the \( \Op{7} \) operator, shown in the top panels of Fig.~\ref{fig:c7c10Compare}.
	\( \Op{7} \) is spin-dependent and velocity-suppressed.
	It is unclear however whether such an operator can be the dominant operator in a full theory.
	It can occur in the presence of a parity-violating vector interaction, but this also comes in a linear combination with \( \Op{9} \) operator~\cite{Dent:2015zpa}, which does not couple to the total nucleus spin, and therefore is more readily accessible at direct detection experiments.

	Results for (\(\Op{ 10 } = i \hermS{\N} \cdot \frac{\hermq}{\mN}\)) are also shown in Fig.~\ref{fig:c7c10Compare}.
	This operator shares some similarity to the \ac{SD} constraints in Fig.~\ref{fig:c1c4Compare} and Fig.~\ref{fig:triple4all}, in that the neutrino constraints are competitive with, but not dominant over other \ac{DD} constraints.
	Finally, we show results for (\(\Op{ 11 } = i \hermS{\chi} \cdot \frac{\hermq}{\mN}\)) in Fig.~\ref{fig:double11best}, whose spin-independent, momentum-suppressed cross section leads \ac{DD} experiments to dominate constraints.

	A number of operators, including \( \Op{ 10 } \) and \( \Op{ 11 } \), show a higher-likelihood (brighter colour) region just below the exclusion lines.
	These are driven by xenon experiments, and LZ in particular, which see downward fluctuations at lower recoil energies with respect to the background model.
	This results in a slightly better fit for dark matter models that predict a rise in recoil rates with energy.
	These higher-likelihood regions are not significant, however, as can be seen from the profile likelihood ratio colour scale.

	To explain the relative power of solar versus direct detection data, we may look at how each of these operators couple to nuclei.
	As stated above, \( \Op{7} \) and \( \Op{10} \) depend on the nuclear spin, so are more difficult to reach by experiments with no spin-dependent sensitivity.
	\( \Op{10} \) is the projection of the spin onto the momentum transfer, while \( \Op{7} \) projects the spin perpendicular to that direction~\cite{Fitzpatrick:2012ib}.
	Scattering in the Sun for these operators is thus dominated by elements with unpaired nucleons, namely hydrogen, and (specifically for \( \Op{10} \) at high \( m_\chi \)) \( ^{14}\text{Ne} \).
	\( \Op{7} \) scales as \( {(\hermv)}^2 \), so it benefits from the higher DM velocities within the Sun's gravitational well.
	This leads to neutrino telescopes being more sensitive in the case of annihilation to \( \WW \) and \( \tautau \) in the high-mass IceCube region, as well as at the lowest masses considered for all annihilation channels, thanks to \superk[].

	At high masses, the \( q^2 \) momentum-dependence of \( \Op{10} \) leads to a suppression in the Sun, since the large ratio between DM and target mass leads to a low average momentum transfer.
	Solar capture remains competitive despite this, thanks to the insensitivity of large DD experiments to spin-dependent couplings.

	In contrast \( \Op{11} \) acts as a \ac{SI}-like interaction but with a \( q^2 \) scaling in contrast with the constant scaling of \( \Op{1} \).
	As with \( \Op{10} \), this leads to a kinematic mismatch between DM and target nuclei in the Sun, in the \( \sim 100 \) GeV range where solar constraints dominate.
	This remains true even despite the fact that \( \Op{11} \) dominantly scatters with iron (see e.g. Fig. 6 of~\cite{Catena:2015iea}).


\begin{figure*}[htb]
		\centering
		\begin{minipage}[t]{0.49\textwidth}
			\includegraphics[width=\textwidth]{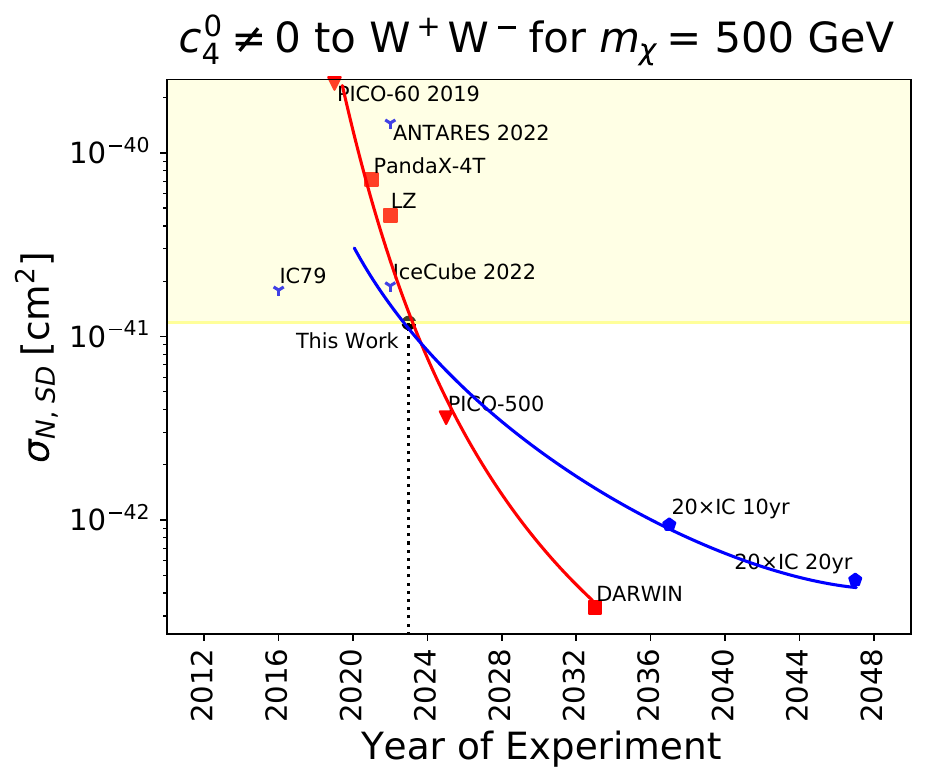}
		\end{minipage}
		\begin{minipage}[t]{0.49\textwidth}
			\includegraphics[width=\textwidth]{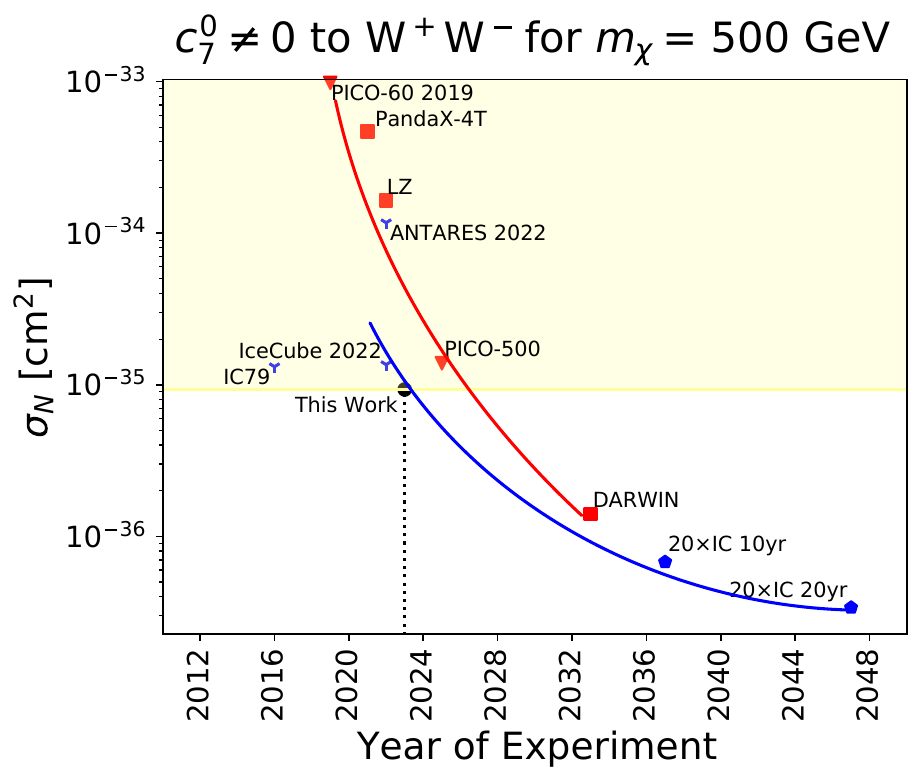}
		\end{minipage}
		\caption[]{Here we plot the 90\% CL from each experiment at a fixed \ac{DM} mass of \( m_\text{dm} = \qty{500}{\GeV} \), as a function of year of the experiment.
		Included are projections from \darwin[c] and \picoFive[c] from \ac{GAMBIT}, and a projection of a future IceCube experiment at twenty times the volume of the current detector.
		The red (\ac{DD}) and blue (solar neutrino) lines are drawn to guide the eye.
		The vertical dotted line indicates the current year and coincides with the point that indicates the current total 90\% CL labeled \emph{This Work}.
		The horizontal light yellow line and region indicates the 90\% CL from the total.
		Left: \( \couple[0]{4} \).
		Right: \( \couple[0]{7} \).
		In comparing the left-hand and right-hand plots, we can see that even for couplings like \( \couple[0]{4} \) where the current solar neutrino experiments beat current \ac{DD} experiments, the projections indicate that in future this will not hold.
		In contrast, for a coupling such as \( \couple[0]{7} \) projections indicate that there is a small window of opportunity for future neutrino experiments to continue to be competitive with future \ac{DD} experiments.
		The remaining operator couplings can be seen in Appendix~\ref{appendix:projection}.}%
		\label{fig:c4c7WWfuture}
	\end{figure*}

	Next, we turn to projections for the sensitivity of future \ac{DD} and neutrino experiments.
	We focus on \( m_\text{dm} = \qty{500}{\GeV} \), where IceCube (2022)'s data are the most constraining, and the \( \WW \) final state.
	A selection of 90\% exclusions from \ac{DD} experiments and neutrino telescopes is shown in the left panel of Fig.~\ref{fig:c4c7WWfuture} for the spin-dependent equivalent coupling \( \couple{4} \).
	In addition to current experiments, we use \ac{GAMBIT} to produce future projections for \darwin[c] and \picoFive[c], which we place at 2033~\cite{Cooley:2022ufh} and 2025~\cite{Akerib:2022ort} respectively.
	These sensitivities are implemented in DDCalc.
	These projections have received more recent updates since the previous DDCalc release, but we include them as they still remain relevant to represent the sensitivity of future \ac{DD} experiments.
	For DARWIN, DDCalc assumes 200 ton-years of exposure~\cite{Schumann:2015cpa}, while PICO-500, DDCalc uses the PICO-500 projection assumptions detailed in from~\cite{Fallows:2017}, namely \qty{250}{\L} fiducial volume of \( \text{C}_3\text{F}_8 \) with 6 live months at \qty{3.2}{\keV} threshold, and 12 months at \qty{10}{\keV}.

	In Fig.~\ref{fig:c4c7WWfuture} we also provide an estimate of a future combined neutrino telescope sensitivity, where we assume that sensitivity simply scales with statistics.
	We estimate the total volume to be twenty times the current IceCube volume, the approximate combined volume of upcoming large-scale optical neutrino telescopes IceCube-Gen2~\cite{IceCube-Gen2:2020qha}, KM3NeT~\cite{KM3Net:2016zxf}, the Baikal-GVD~\cite{Baikal-GVD:2018isr}, P-ONE~\cite{P-ONE:2020ljt}, and TRIDENT~\cite{Ye:2022vbk}.
	The neutrino telescope projection is taken from a reference value for the current exclusion provided by IceCube 2022, and scaling it with the square root of the scaled volume and duration of experiment:
	\begin{equation}
		\text{Projection} = \text{IC2022} \, {\left(\frac{V}{V_\text{IC2022}} \frac{T}{T_\text{IC2022}}\right)}^{-\frac{1}{2}} \, .
		\label{eqn:nuProj}
	\end{equation}
	The left panel of Fig.~\ref{fig:c4c7WWfuture} shows that the excluding power on \ac{NREO} \ac{DM} from future underground \ac{DD} experiments is very likely to outpace that from neutrino telescopes.
	Only in a tight \ac{DM} mass window for \( \couple{7} \), as seen in right panel of Fig.~\ref{fig:c4c7WWfuture} does a future projection from neutrino telescopes appear competitive.
	We show the remaining excluding power plots for each of the 14 operators in Appendix~\ref{appendix:projection}.
	Future solar neutrino analyses only outperform future \ac{DD} experiments for \( \couple[0]{7} \) (and perhaps remaining competitive in the case of \( \couple[0]{14} \), see Fig.~\ref{fig:c14c15WWfuture}) in a small mass region around \( m_\text{DM} = \qty{500}{\GeV} \).

\section{Conclusions}\label{sec:Conclusions}

	We have presented a global fit of \ac{NREO} interactions between dark matter and \ac{SM} particles, combining \ac{DD} experiments and neutrino telescope observations, including the most recent constraints from \ic[] and \antares[].
	While these methods remain complementary for some operators, in future \ac{DD} experiments will remain dominant in most scenarios.

	The region of parameter space in which \ac{DM} annihilation to neutrino final products exceed sensitivities of constructed \ac{DD} experiments is thus swiftly closing.
	The neutrino fog~\cite{Monroe:2007xp,Strigari:2009bq,Billard:2013qya,OHare:2021utq}, which we have not modeled in our projections will become an important background over the next decade, potentially slowing the direct detection side of the race.
	Simultaneously, solar atmospheric neutrinos (the ``neutrino mist'')~\cite{Ng:2017aur,Arguelles:2017eao} will begin to pose a challenge to searches for new physics in high-energy neutrinos from the Sun.
	Both methods thus may well remain competitive and complementary, and, as we have shown, combining the two in a self-consistent and systematic way yields more than the sum of their parts.

\acknowledgments{}
	We thank Ben Farmer, Felix Kahlhoefer and Sebastian Wild for valuable input. ACV and NPAK are supported by the Arthur B. McDonald Canadian Astroparticle Physics Research Institute, NSERC and the Ontario Ministry of Colleges and Universities, with equipment funded by the Canada Foundation for Innovation and the Ontario Government.
	Research at Perimeter Institute is supported by the Government of Canada through the Department of Innovation, Science, and Economic Development, and by the Province of Ontario.
	PS is supported by Australian Research Council Discovery Project DP220100643.

\FloatBarrier
\bibliographystyle{JHEP_pat}
\bibliography{myBib}

\onecolumngrid
\appendix
\section{All Scans}\label{appendix:allscans}
	\begin{figure*}[htb]
		\includegraphics{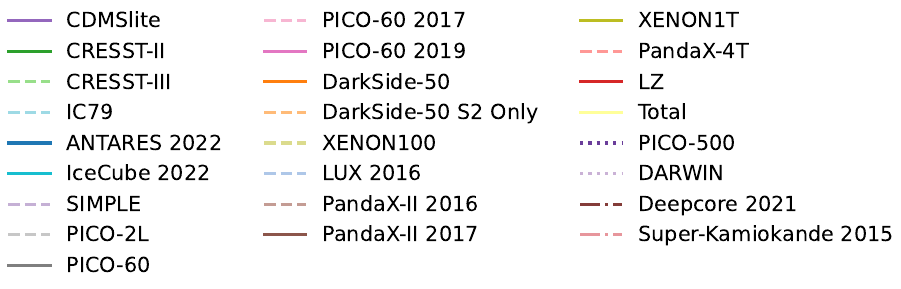}
		\caption{The legend of experiment likelihoods for all plots, rendered separately so that the plots themselves can be viewed more clearly on their own.}%
		\label{fig:legend}
	\end{figure*}
	\begin{figure*}[htb]
		\centering
		\centerline{\includegraphics[width=\paperwidth]{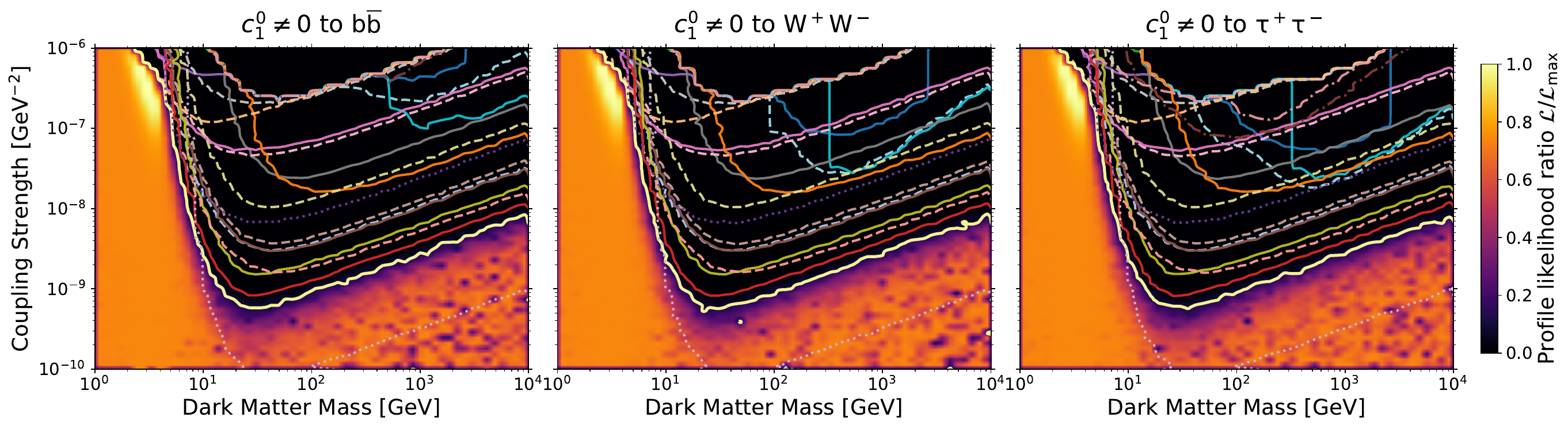}}
		\caption[Profile likelihood of \({\couple[0]{1}}\)]{Profile likelihood ratios in the plane of dark matter mass and coupling strength for \( \couple[0]{1} \).
		Left: annihilation channel to \( \bb \), center: \( \WW \), and right: \( \tautau \).
		The lines indicate the 90\% CL contours of the total and each experiment's contribution to the total.
		The coloured histogram represents the profile likelihood ratio \( \mathcal{L} / \mathcal{L}_\text{max} \) of the total.
		The legend for each experiment's contour can be seen in Fig.~\ref{fig:legend}.
		The IC79 line \emph{does not} contribute to the total above \qty{300}{\GeV} as it has overlap with IceCube (2022).}%
		\label{fig:triple1all}
	\end{figure*}
	\begin{figure*}[htb]
		\centering
		\centerline{\includegraphics[width=\paperwidth]{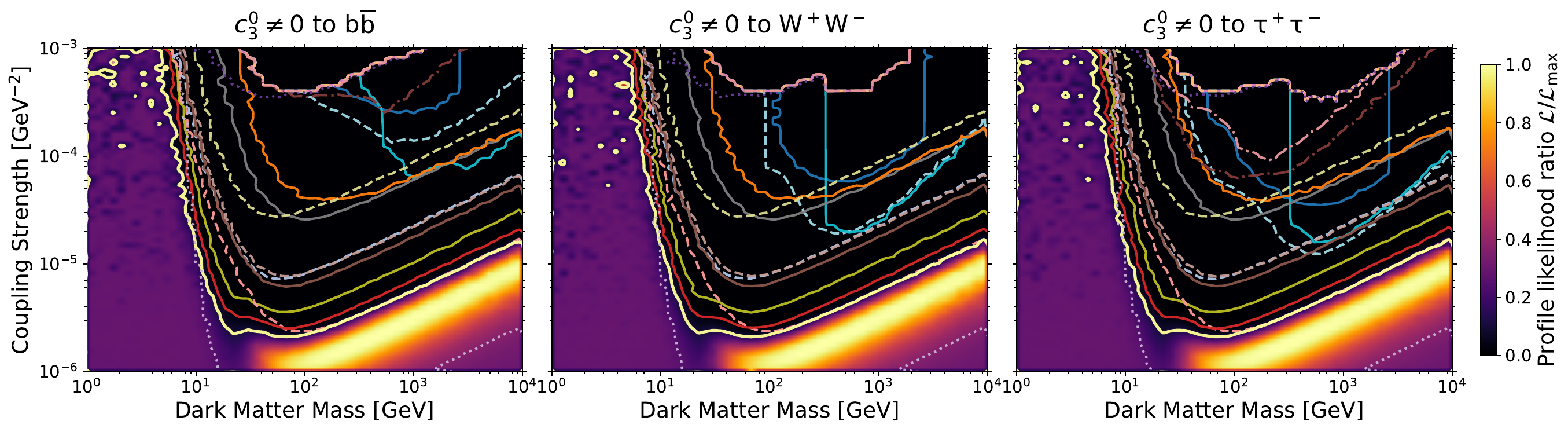}}
		\caption[Profile likelihood of \({\couple[0]{3}}\)]{Same as Fig.~\ref{fig:triple1all}, but for \(\couple[0]{3}\).}%
		\label{fig:triple3all}
	\end{figure*}
	\begin{figure*}[htb]
		\centering
		\centerline{\includegraphics[width=\paperwidth]{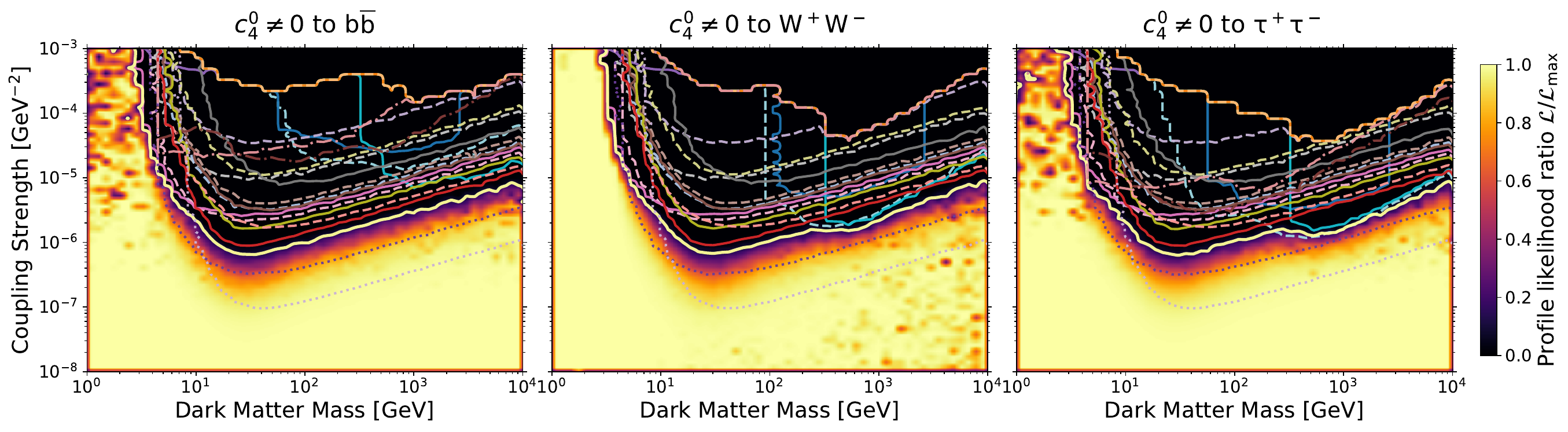}}
		\caption[Profile likelihood of \({\couple[0]{4}}\)]{Same as Fig.~\ref{fig:triple1all}, but for \(\couple[0]{4}\).}%
		\label{fig:triple4all}
	\end{figure*}
	\begin{figure*}[htb]
		\centering
		\centerline{\includegraphics[width=\paperwidth]{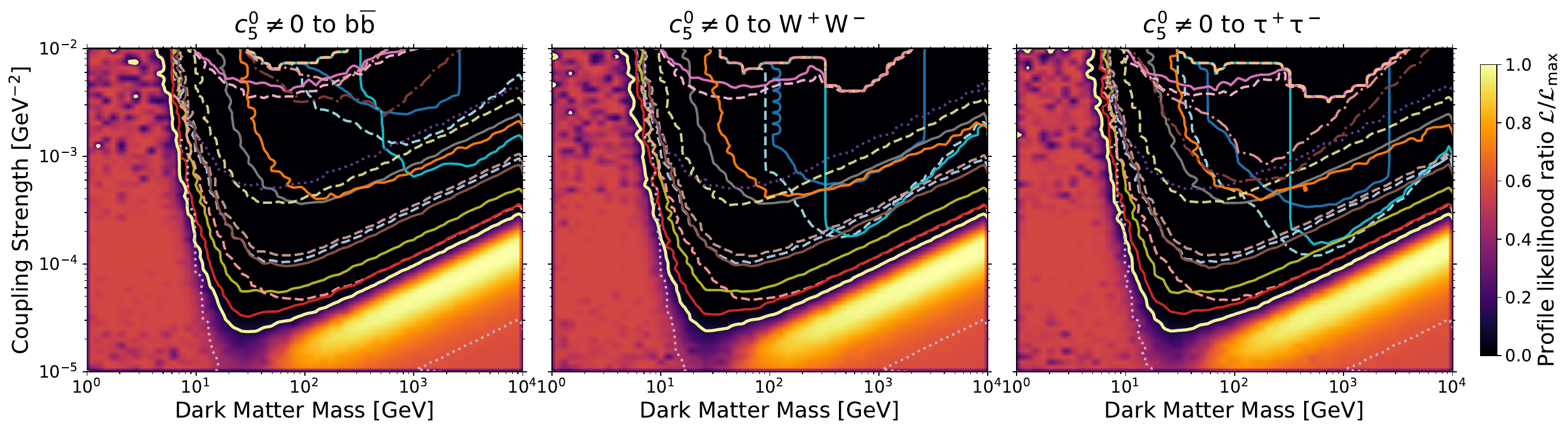}}
		\caption[Profile likelihood of \({\couple[0]{5}}\)]{Same as Fig.~\ref{fig:triple1all}, but for \(\couple[0]{5}\).}%
		\label{fig:triple5all}
	\end{figure*}
	\begin{figure*}[htb]
		\centering
		\centerline{\includegraphics[width=\paperwidth]{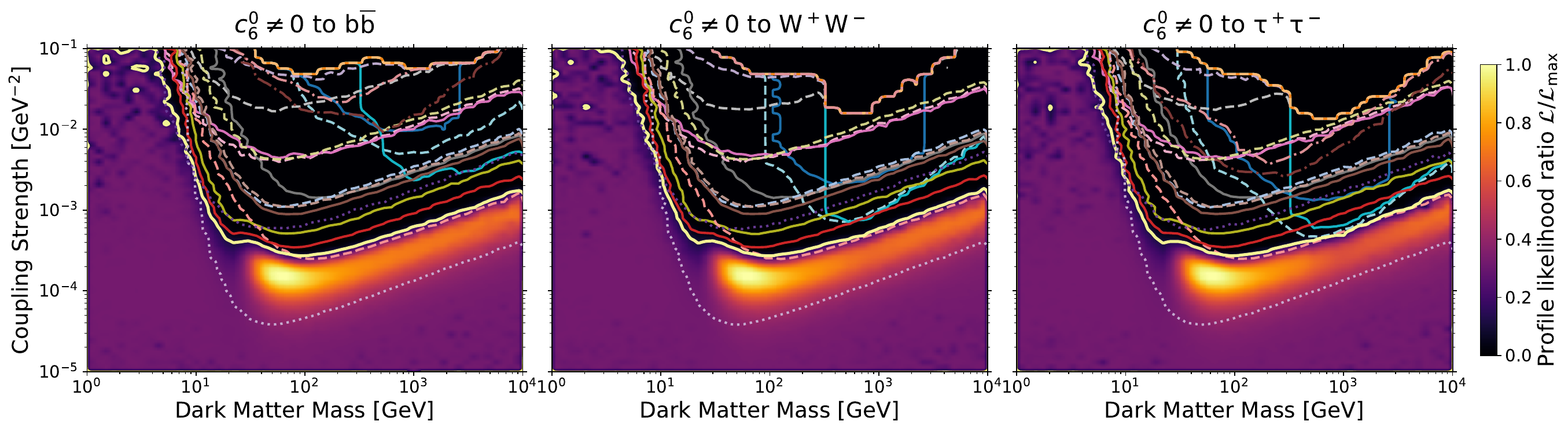}}
		\caption[Profile likelihood of \({\couple[0]{6}}\)]{Same as Fig.~\ref{fig:triple1all}, but for \(\couple[0]{6}\).}%
		\label{fig:triple6all}
	\end{figure*}
	\begin{figure*}[htb]
		\centering
		\centerline{\includegraphics[width=\paperwidth]{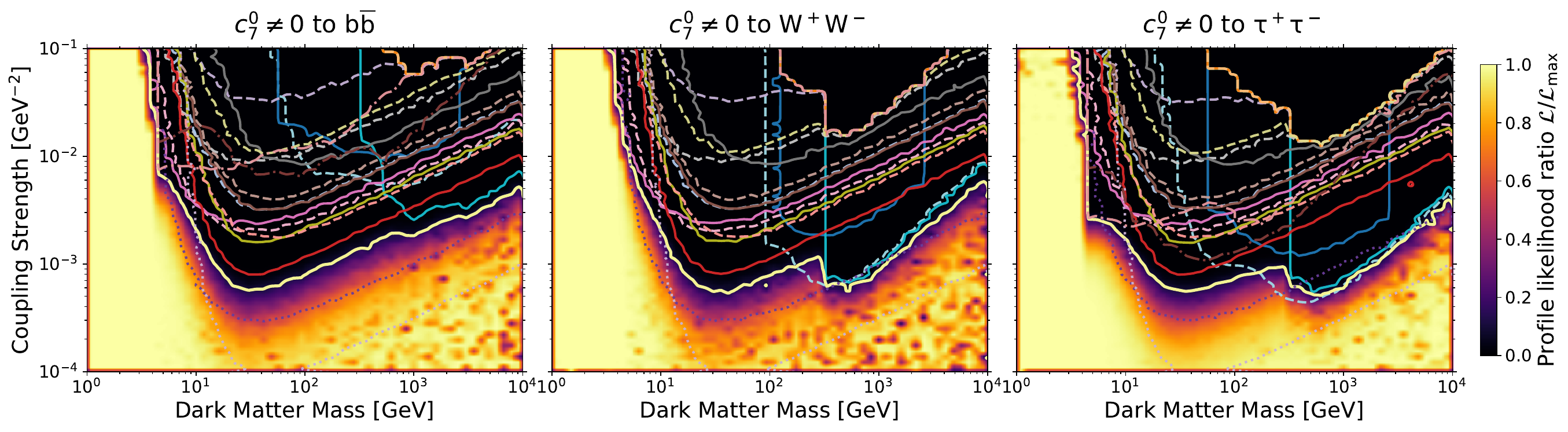}}
		\caption[Profile likelihood of \({\couple[0]{7}}\)]{Same as Fig.~\ref{fig:triple1all}, but for \(\couple[0]{7}\).}%
		\label{fig:triple7all}
	\end{figure*}
	\begin{figure*}[htb]
		\centering
		\centerline{\includegraphics[width=\paperwidth]{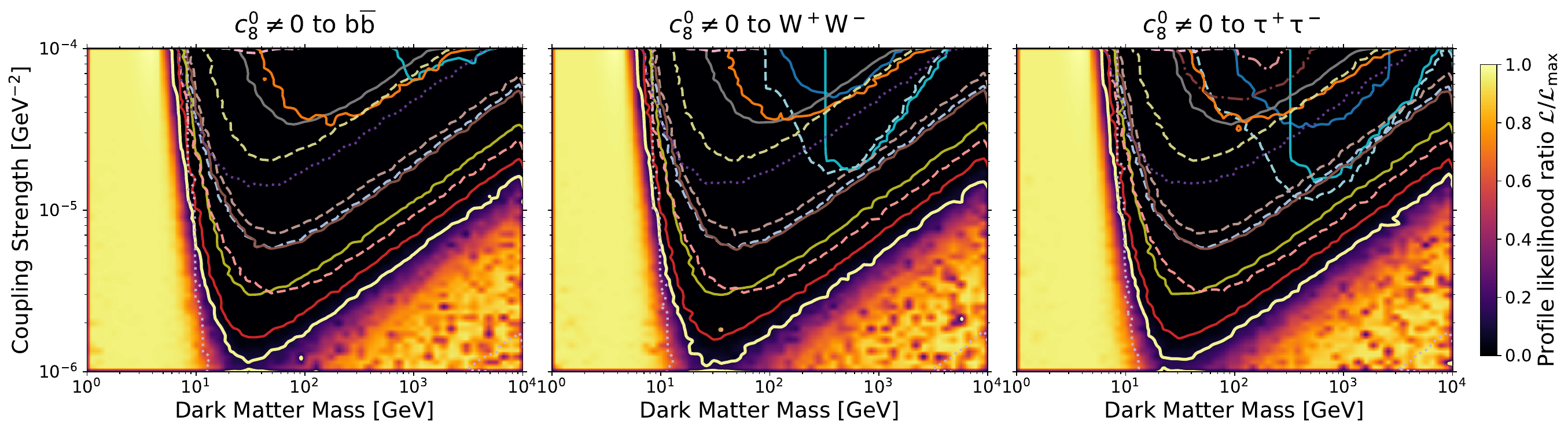}}
		\caption[Profile likelihood of \({\couple[0]{8}}\)]{Same as Fig.~\ref{fig:triple1all}, but for \(\couple[0]{8}\).}%
		\label{fig:triple8all}
	\end{figure*}
	\begin{figure*}[htb]
		\centering
		\centerline{\includegraphics[width=\paperwidth]{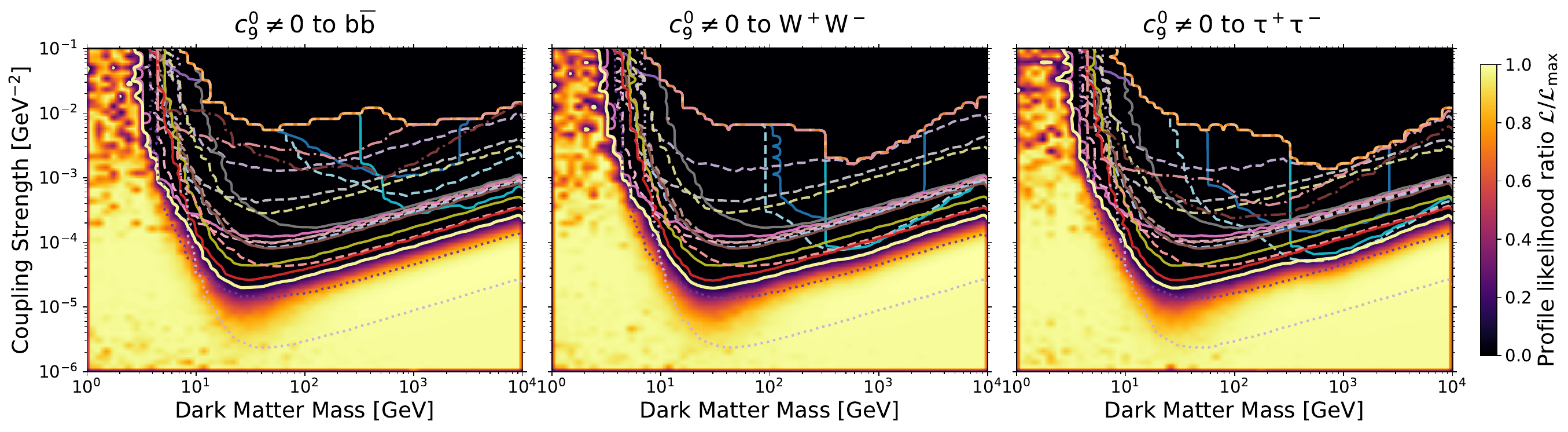}}
		\caption[Profile likelihood of \({\couple[0]{9}}\)]{Same as Fig.~\ref{fig:triple1all}, but for \(\couple[0]{9}\).}%
		\label{fig:triple9all}
	\end{figure*}
	\begin{figure*}[htb]
		\centering
		\centerline{\includegraphics[width=\paperwidth]{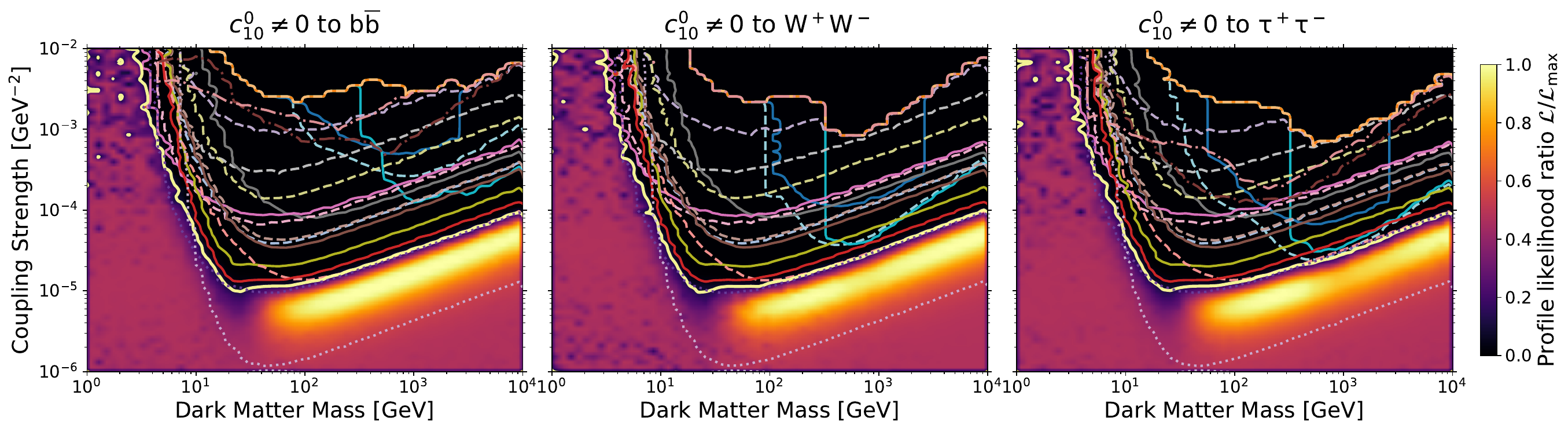}}
		\caption[Profile likelihood of \({\couple[0]{10}}\)]{Same as Fig.~\ref{fig:triple1all}, but for \(\couple[0]{10}\).}%
		\label{fig:triple10all}
	\end{figure*}
	\begin{figure*}[htb]
		\centering
		\centerline{\includegraphics[width=\paperwidth]{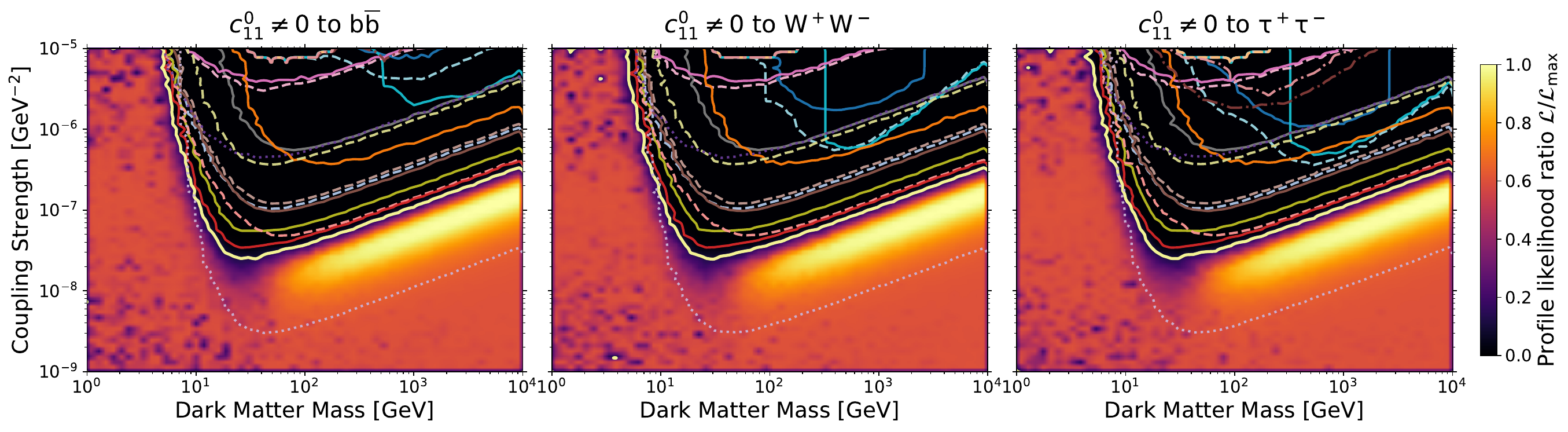}}
		\caption[Profile likelihood of \({\couple[0]{11}}\)]{Same as Fig.~\ref{fig:triple1all}, but for \(\couple[0]{11}\).}%
		\label{fig:triple11all}
	\end{figure*}
	\begin{figure*}[htb]
		\centering
		\centerline{\includegraphics[width=\paperwidth]{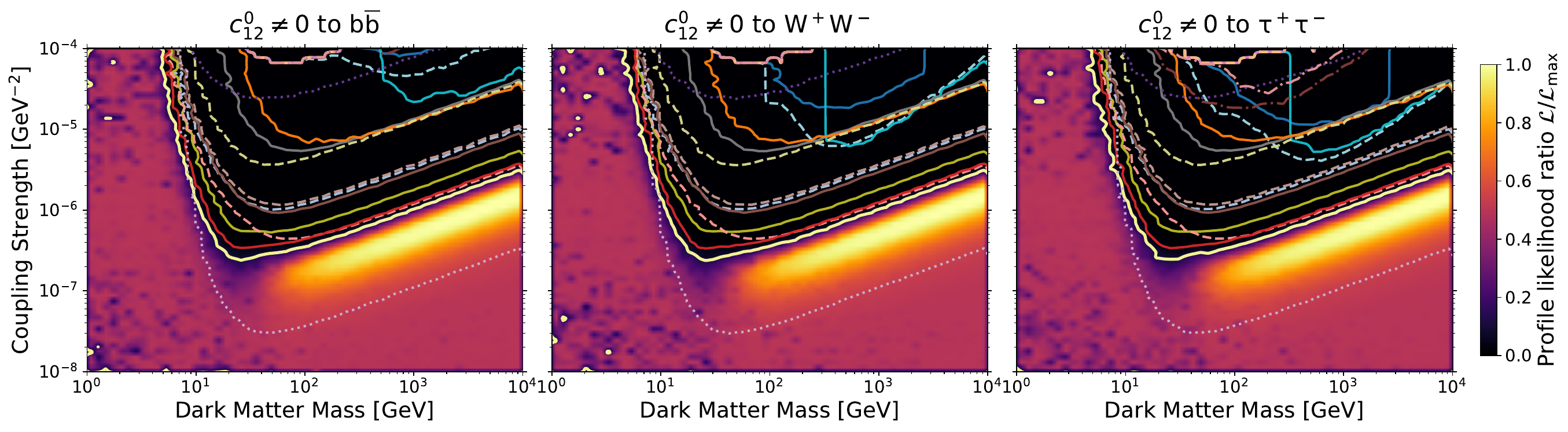}}
		\caption[Profile likelihood of \({\couple[0]{12}}\)]{Same as Fig.~\ref{fig:triple1all}, but for \(\couple[0]{12}\).}%
		\label{fig:triple12all}
	\end{figure*}
	\begin{figure*}[htb]
		\centering
		\centerline{\includegraphics[width=\paperwidth]{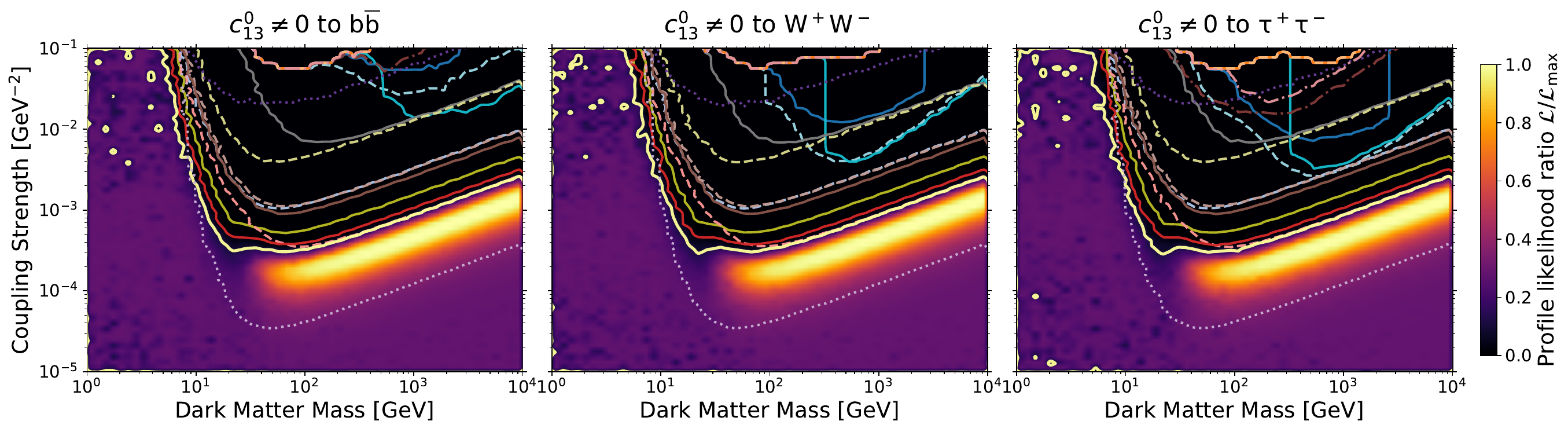}}
		\caption[Profile likelihood of \({\couple[0]{13}}\)]{Same as Fig.~\ref{fig:triple1all}, but for \(\couple[0]{13}\).}%
		\label{fig:triple13all}
	\end{figure*}
	\begin{figure*}[htb]
		\centering
		\centerline{\includegraphics[width=\paperwidth]{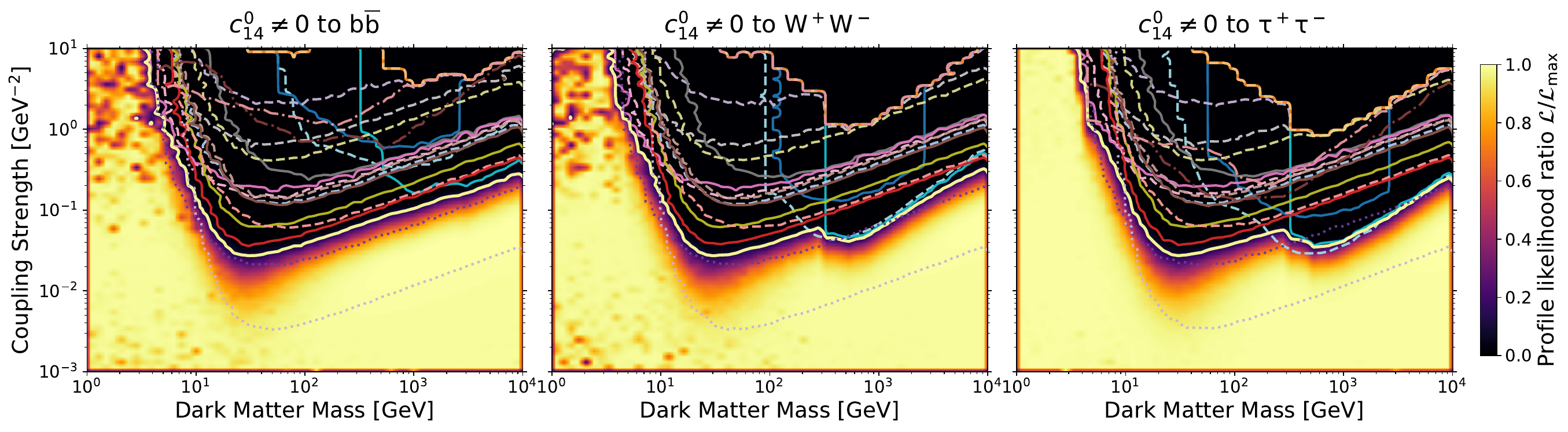}}
		\caption[Profile likelihood of \({\couple[0]{14}}\)]{Same as Fig.~\ref{fig:triple1all}, but for \(\couple[0]{14}\).}%
		\label{fig:triple14all}
	\end{figure*}
	\begin{figure*}[htb]
		\centering
		\centerline{\includegraphics[width=\paperwidth]{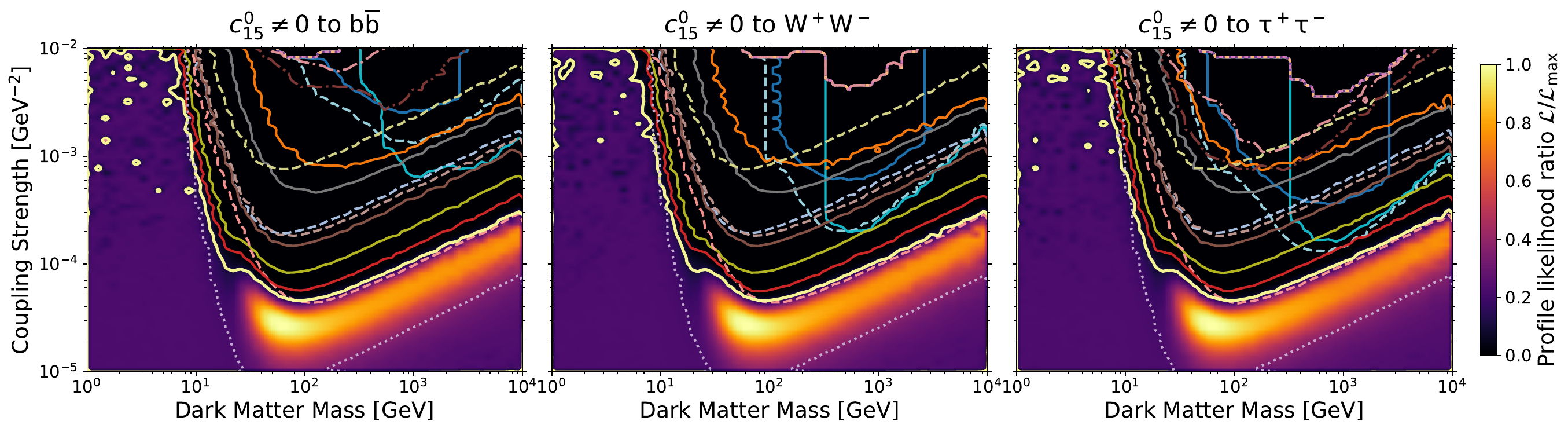}}
		\caption[Profile likelihood of \({\couple[0]{15}}\)]{Same as Fig.~\ref{fig:triple1all}, but for \(\couple[0]{15}\).}%
		\label{fig:triple15all}
	\end{figure*}

\FloatBarrier
\section{Projection Plots}\label{appendix:projection}
	\begin{figure*}[htb]
		\centering
		\begin{minipage}[t]{0.49\textwidth}
			\includegraphics[width=\textwidth]{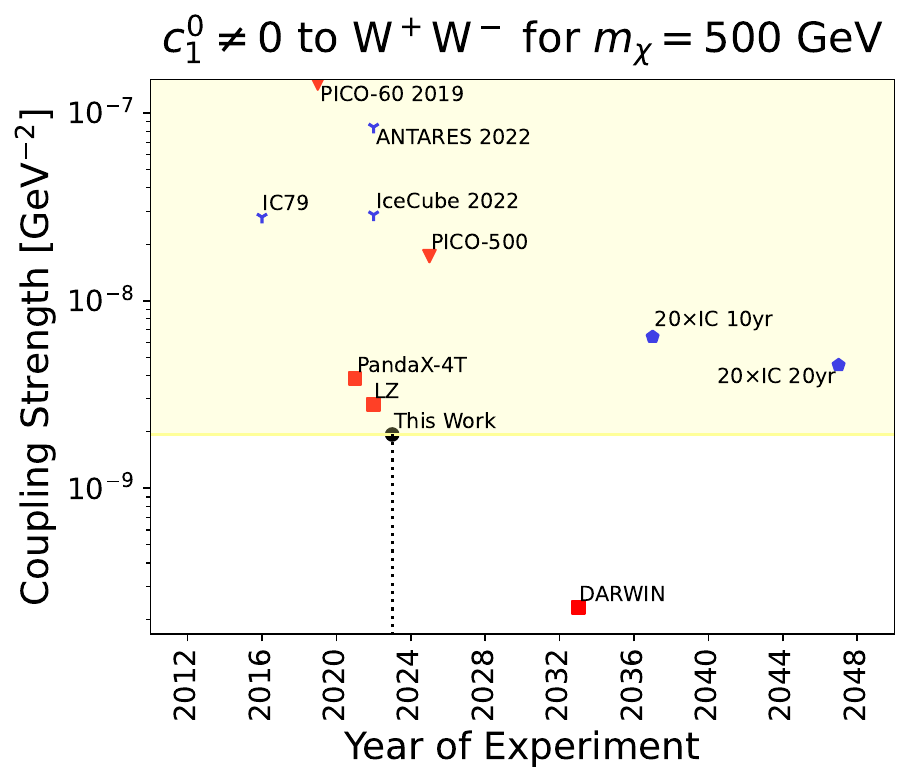}
		\end{minipage}
		\begin{minipage}[t]{0.49\textwidth}
			\includegraphics[width=\textwidth]{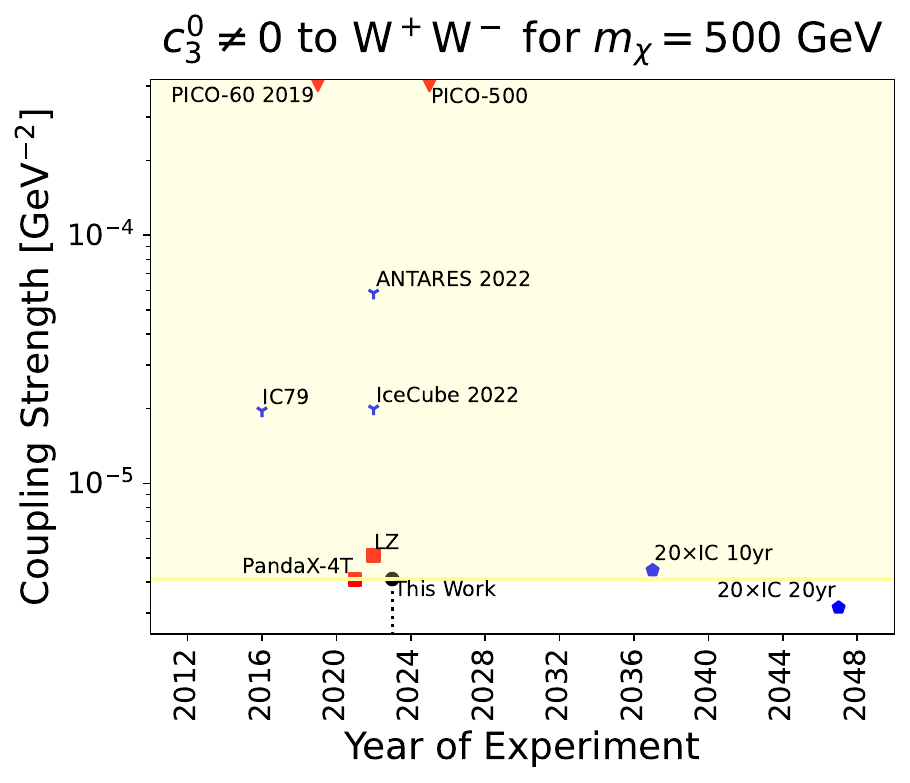}
		\end{minipage}
		\caption[]{Here we plot the 90\% CL from each experiment at a fixed \ac{DM} mass of \( m_\text{dm} = \qty{500}{\GeV} \), as a function of year of the experiment.
		Included are projections from \darwin[c] and \picoFive[c] from \ac{GAMBIT}, and a projection of a future IceCube experiment at twenty times the volume of the current detector.
		The vertical black dotted line indicates the current year and coincides with the point that indicates the current total 90\% CL labeled \emph{This Work}.
		The horizontal light yellow line and region indicates the 90\% CL from the total.
		Left: \( \couple[0]{1} \).
		Right: \( \couple[0]{3} \).}%
		\label{fig:c1c3WWfuture}
	\end{figure*}
	\begin{figure*}[htb]
		\centering
		\begin{minipage}[t]{0.49\textwidth}
			\includegraphics[width=\textwidth]{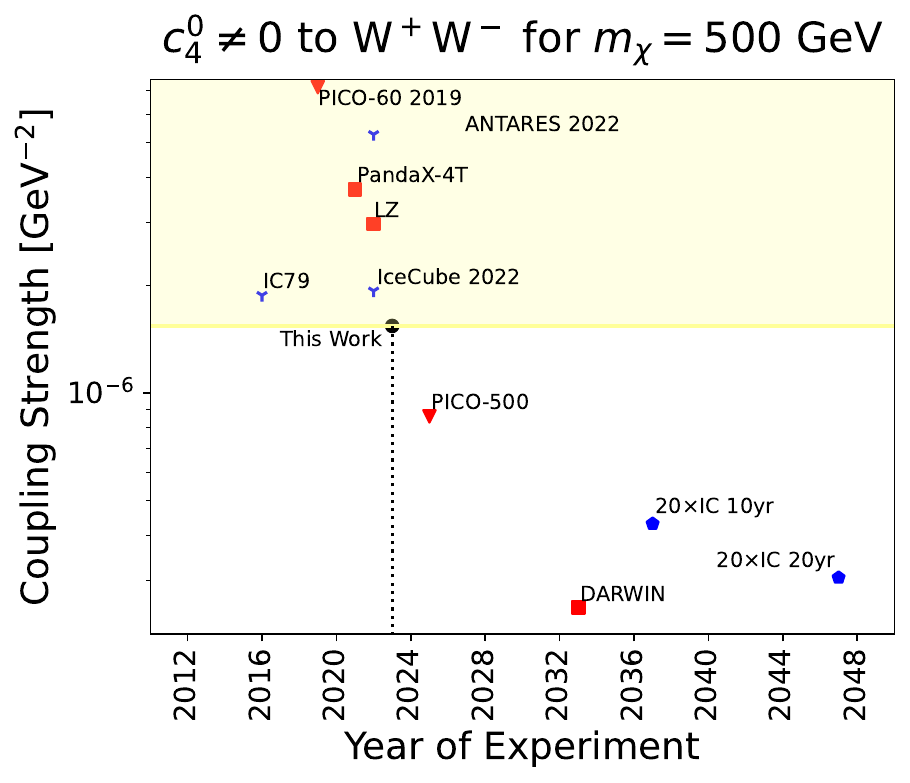}
		\end{minipage}
		\begin{minipage}[t]{0.49\textwidth}
			\includegraphics[width=\textwidth]{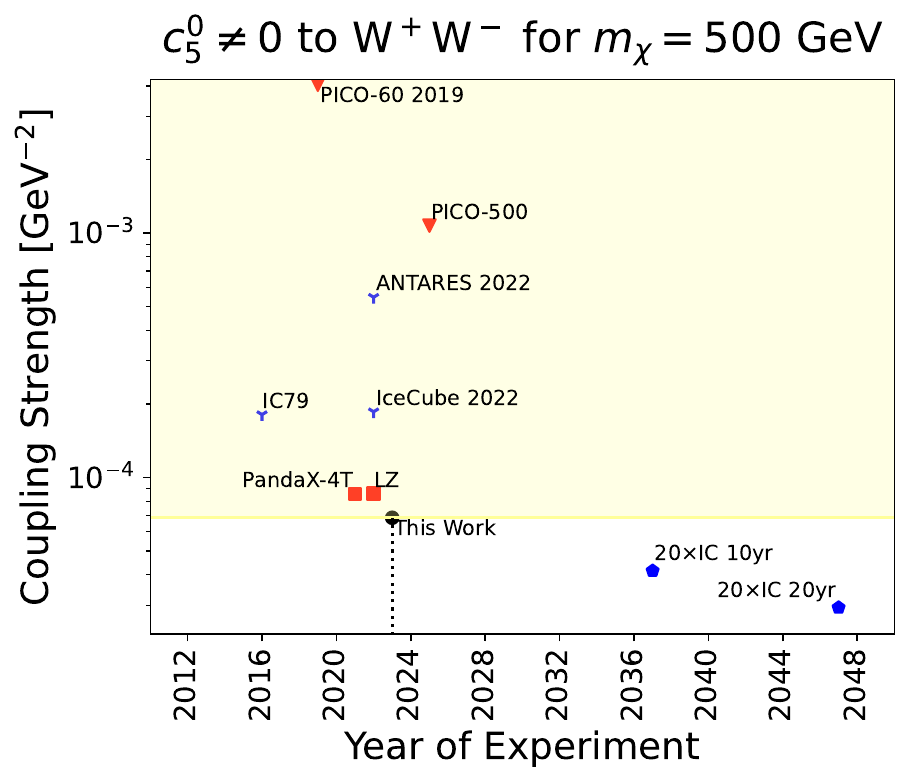}
		\end{minipage}
		\caption[]{The same as Fig.~\ref{fig:c1c3WWfuture}, but for \(\couple[0]{4}\) and \(\couple[0]{5}\).}%
		\label{fig:c4c5WWfuture}
	\end{figure*}
	\begin{figure*}[htb]
		\centering
		\begin{minipage}[t]{0.49\textwidth}
			\includegraphics[width=\textwidth]{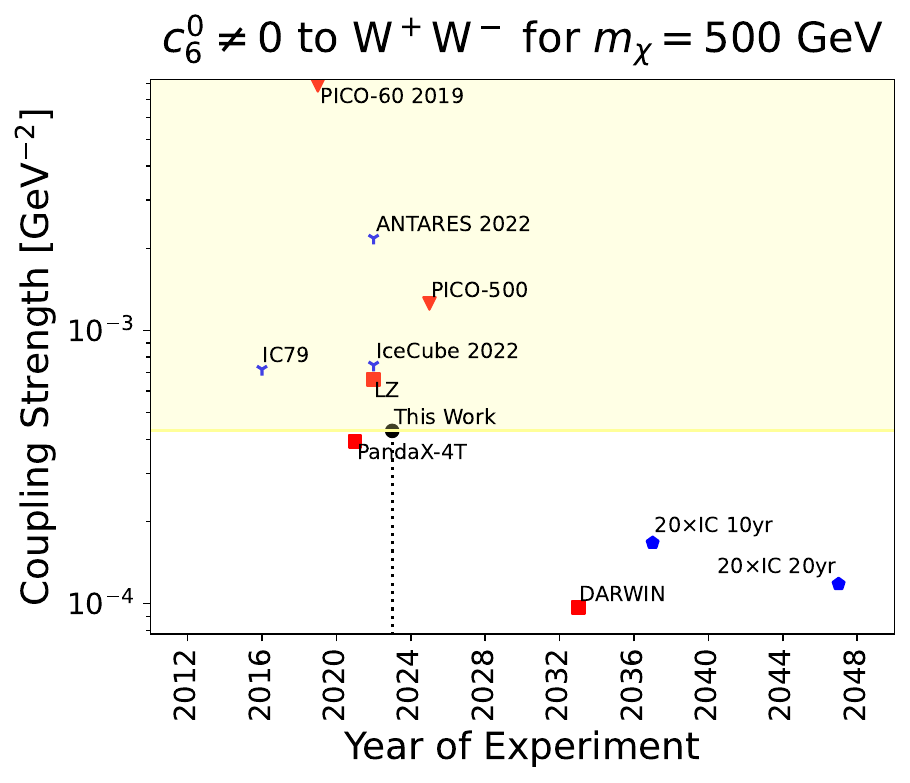}
		\end{minipage}
		\begin{minipage}[t]{0.49\textwidth}
			\includegraphics[width=\textwidth]{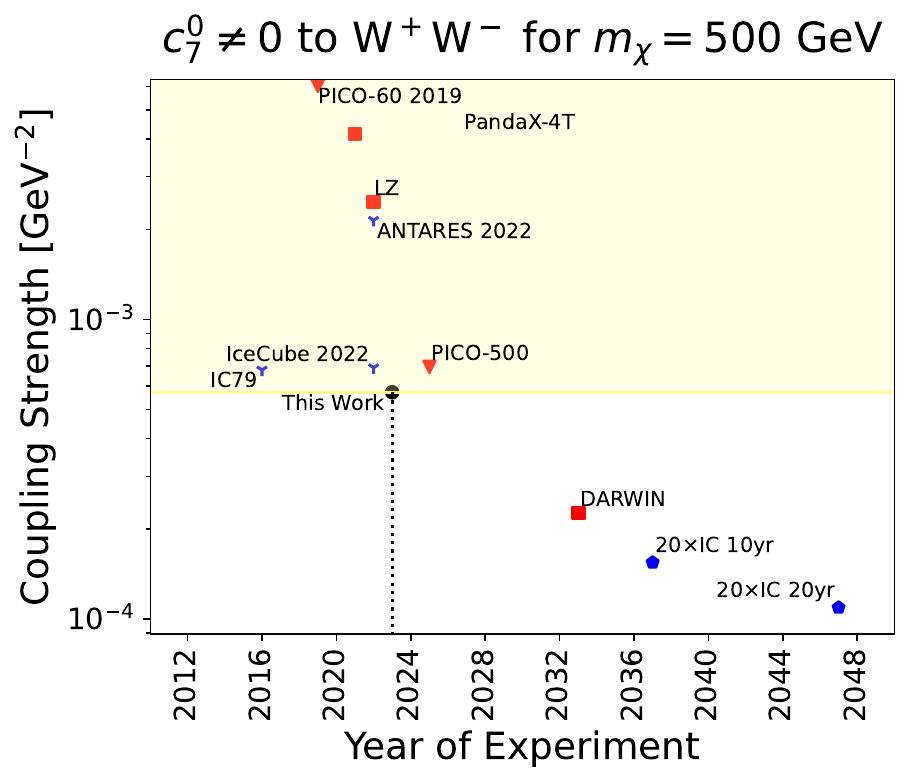}
		\end{minipage}
		\caption[]{The same as Fig.~\ref{fig:c1c3WWfuture}, but for \(\couple[0]{6}\) and \(\couple[0]{7}\).}%
		\label{fig:c6c7WWfuture}
	\end{figure*}
	\begin{figure*}[htb]
		\centering
		\begin{minipage}[t]{0.49\textwidth}
			\includegraphics[width=\textwidth]{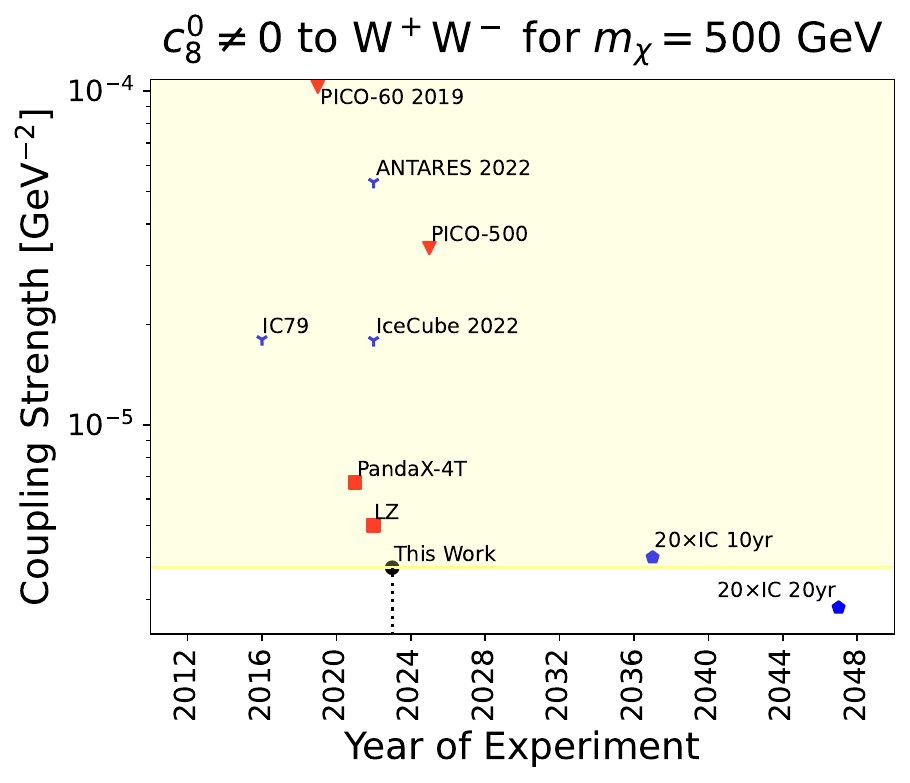}
		\end{minipage}
		\begin{minipage}[t]{0.49\textwidth}
			\includegraphics[width=\textwidth]{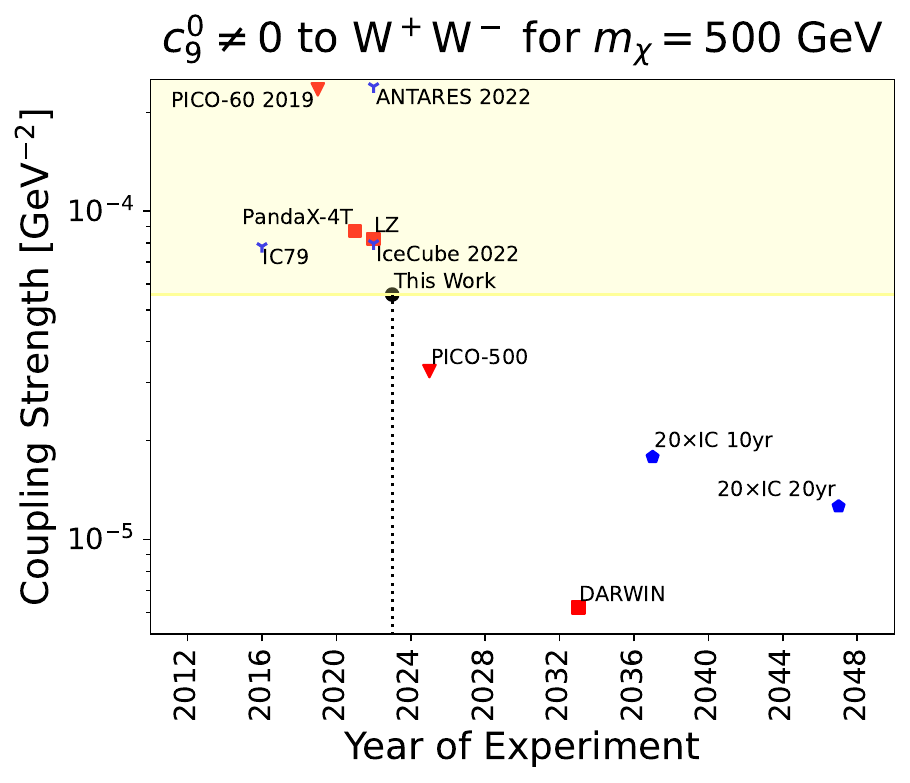}
		\end{minipage}
		\caption[]{The same as Fig.~\ref{fig:c1c3WWfuture}, but for \(\couple[0]{8}\) and \(\couple[0]{9}\).}%
		\label{fig:c8c9WWfuture}
	\end{figure*}
	\begin{figure*}[htb]
		\centering
		\begin{minipage}[t]{0.49\textwidth}
			\includegraphics[width=\textwidth]{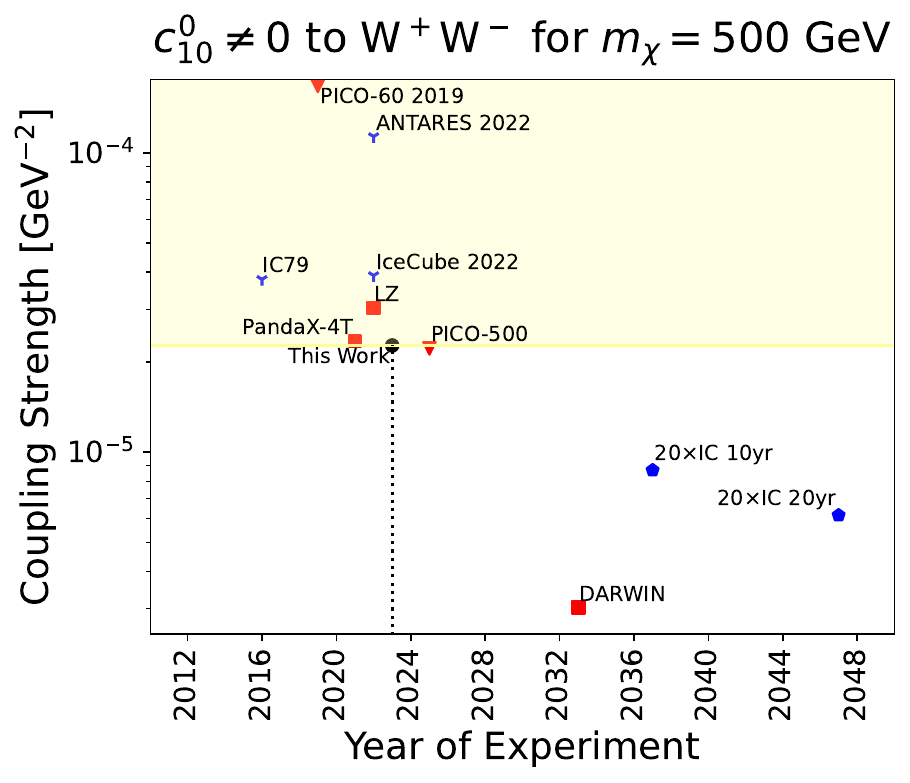}
		\end{minipage}
		\begin{minipage}[t]{0.49\textwidth}
			\includegraphics[width=\textwidth]{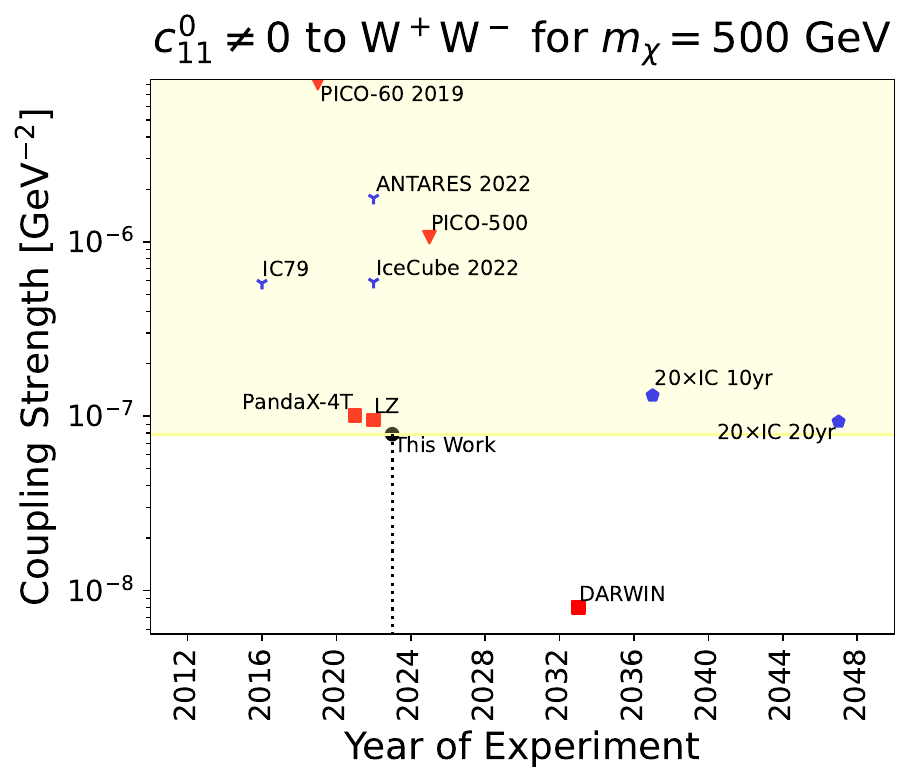}
		\end{minipage}
		\caption[]{The same as Fig.~\ref{fig:c1c3WWfuture}, but for \(\couple[0]{10}\) and \(\couple[0]{11}\).}%
		\label{fig:c10c11WWfuture}
	\end{figure*}
	\begin{figure*}[htb]
		\centering
		\begin{minipage}[t]{0.49\textwidth}
			\includegraphics[width=\textwidth]{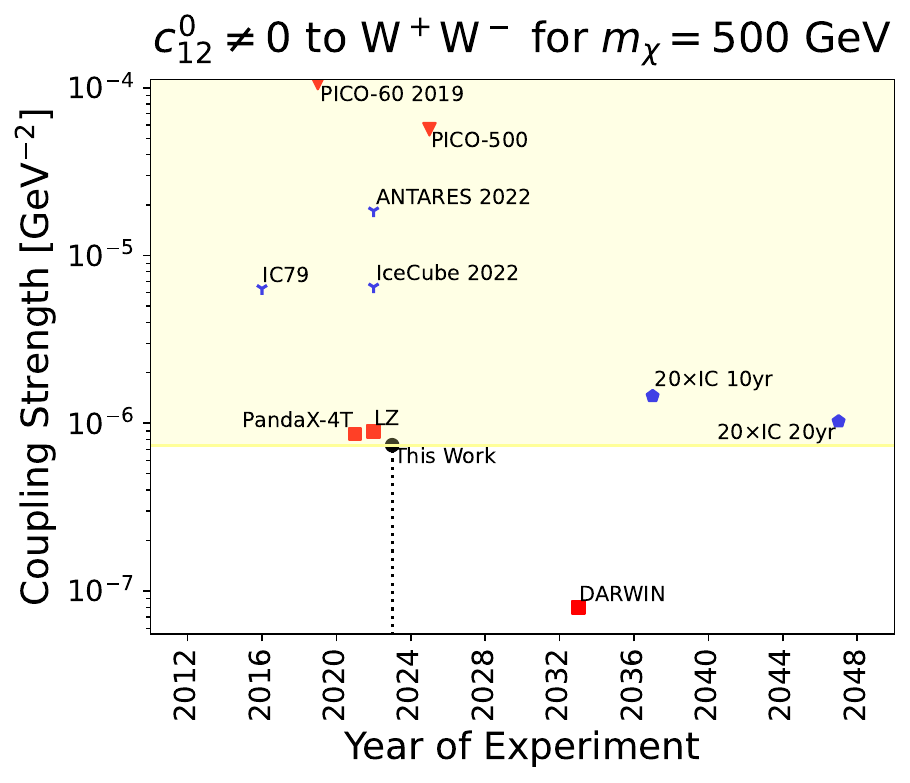}
		\end{minipage}
		\begin{minipage}[t]{0.49\textwidth}
			\includegraphics[width=\textwidth]{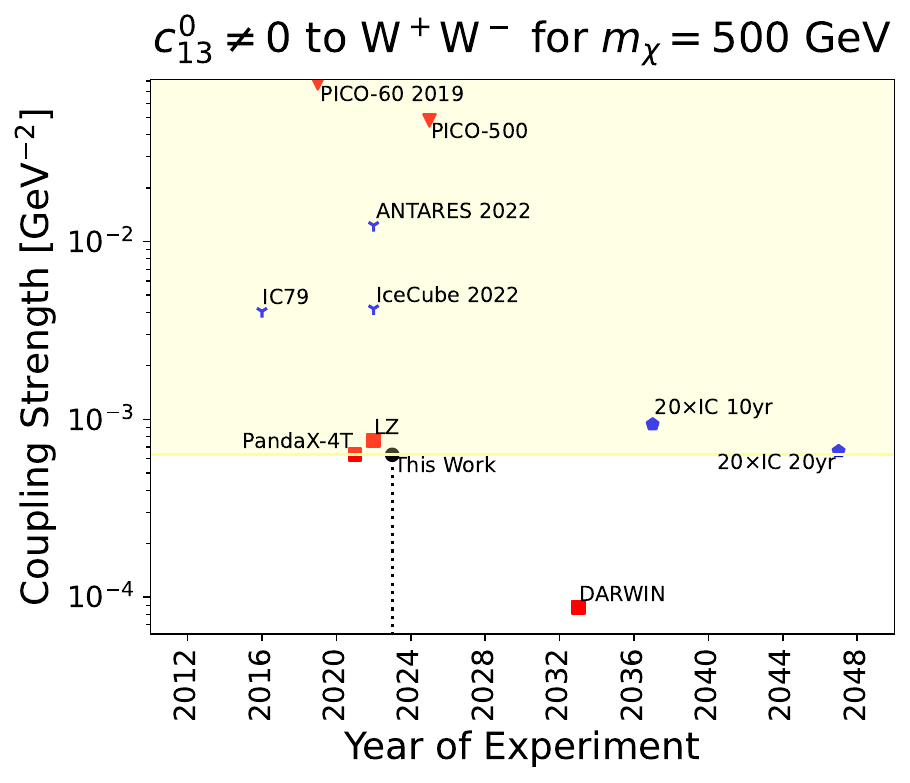}
		\end{minipage}
		\caption[]{The same as Fig.~\ref{fig:c1c3WWfuture}, but for \(\couple[0]{12}\) and \(\couple[0]{13}\).}%
		\label{fig:c12c13WWfuture}
	\end{figure*}
	\begin{figure*}[htb]
		\centering
		\begin{minipage}[t]{0.49\textwidth}
			\includegraphics[width=\textwidth]{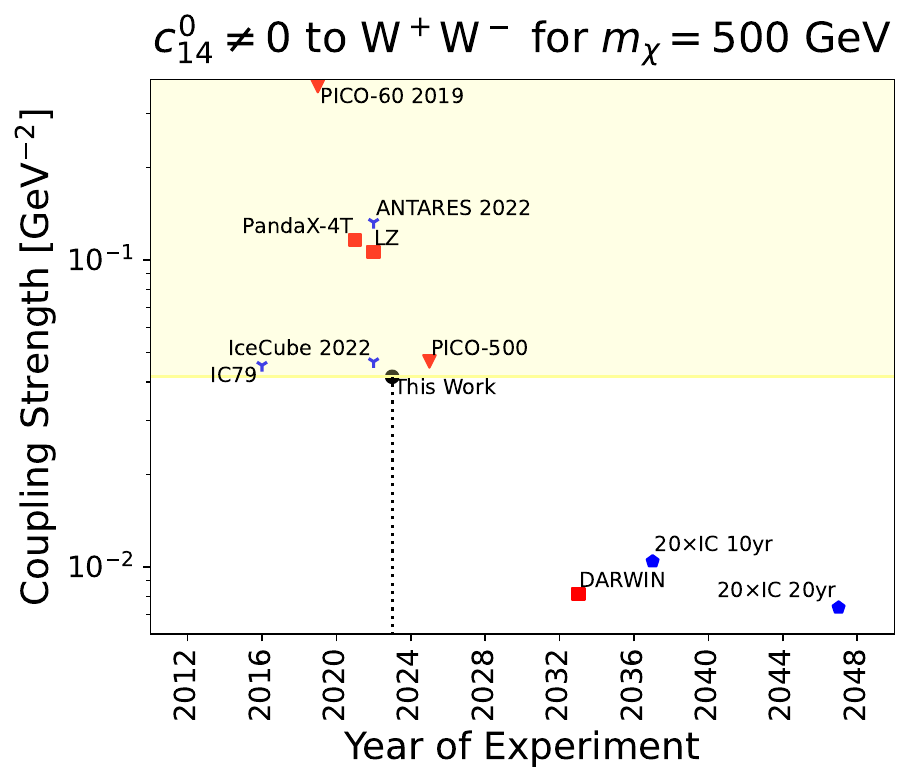}
		\end{minipage}
		\begin{minipage}[t]{0.49\textwidth}
			\includegraphics[width=\textwidth]{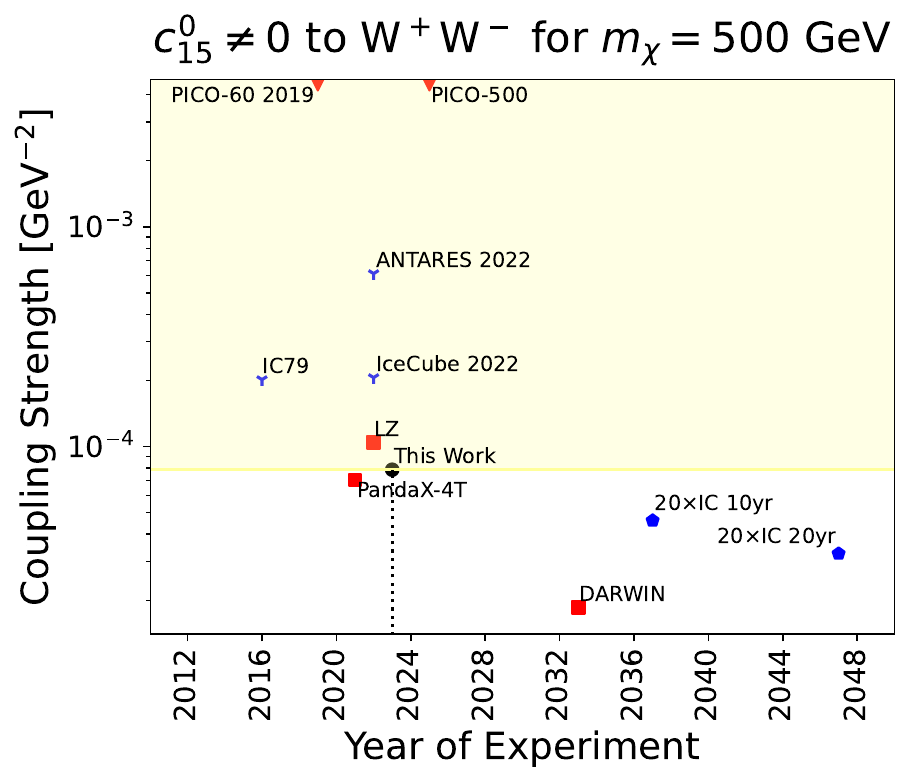}
		\end{minipage}
		\caption[]{The same as Fig.~\ref{fig:c1c3WWfuture}, but for \(\couple[0]{14}\) and \(\couple[0]{15}\).}%
		\label{fig:c14c15WWfuture}
	\end{figure*}

\end{document}